\documentclass[apj]{emulateapj}

\usepackage{natbib}
\usepackage{amsfonts,amsmath,amssymb} 
\usepackage{color}
\usepackage[backref,breaklinks,colorlinks,citecolor=blue]{hyperref} 
\usepackage[all]{hypcap} 

\newcommand{\myrefsec}[1]{\hyperref[#1]{Section~\ref*{#1}}}

\def \Msol{{\rm M}_{\odot}}
\def \Zsol{{\rm Z}_{\odot}}
\def \logm{\log(M/\Msol)}
\def \lya{Ly$\alpha$~}
\def \lyaeq{{\rm Ly}\alpha}
\def \ergscmA{erg~s$^{-1}$~cm$^{2}$~\AA$^{-1}$}

\def \dvis{\Delta v_{{\rm IS}}}
\def \dvlya{\Delta v_{\lyaeq}}
\def \dvlyais{\Delta v_{{\rm Ly}\alpha-{\rm IS}}}

\def \oabund{12+\log({\rm O/H})}
\def \kms{{\rm km~s}^{-1}}
\def \myr{{\rm Myr}}

\def \hbeta{H$\beta$}
\def \halpha{H$\alpha$}

\def \siii{Si{\scriptsize ~II}}
\def \siiii{Si{\scriptsize ~III}}
\def \siiv{Si{\scriptsize ~IV}}
\def \cii{C{\scriptsize ~II}}
\def \ciii{C{\scriptsize ~III}}
\def \civ{C{\scriptsize ~IV}}

\def \oi{O{\scriptsize ~I}}
\def \oii{[O{\scriptsize ~II}]}
\def \oiii{[O{\scriptsize ~III}]}
\def \heii{He{\scriptsize ~II}}
\def \feii{Fe{\scriptsize ~II}}
\def \alii{Al{\scriptsize ~II}}

\def \fev{Fe{\scriptsize ~V}}
\def \feiv{Fe{\scriptsize ~IV}}
\def \niv{N{\scriptsize ~IV}}
\def \nii{[N{\scriptsize ~II}]}
\def \neiii{[N{\scriptsize ~III}]}

\slugcomment{}

\shorttitle{Metallicity of star-forming galaxies at $z \sim 5$}
\shortauthors{A. L. Faisst, et al.}

\begin{document}


\title{Rest-UV absorption lines as metallicity estimator: the metal content of star-forming galaxies at $z \sim 5$}


\author{
A. L. Faisst\altaffilmark{1}$^{,}$\altaffilmark{2},
P. L. Capak\altaffilmark{1}$^{,}$\altaffilmark{2},
I. Davidzon\altaffilmark{3}$^{,}$\altaffilmark{4},
M. Salvato\altaffilmark{5},
C. Laigle\altaffilmark{6},
O. Ilbert\altaffilmark{3},
M. Onodera\altaffilmark{7},
G. Hasinger\altaffilmark{8},
Y. Kakazu\altaffilmark{9},
D. Masters\altaffilmark{1}$^{,}$\altaffilmark{2},
B. Mobasher\altaffilmark{10},
D. Sanders\altaffilmark{8},
J. D. Silverman\altaffilmark{11},
L. Yan\altaffilmark{1},
N. Z. Scoville\altaffilmark{2}
}

\affil{$^{1}$Infrared Processing and Analysis Center, California Institute of Technology, Pasadena, CA 91125, USA}
\affil{$^{2}$Cahill Center for Astronomy and Astrophysics, California Institute of Technology, Pasadena, CA 91125, USA}
\affil{$^{3}$Aix Marseille Universit\'e, CNRS, LAM (Laboratoire d'Astrophysique de Marseille) UMR 7326, 13388, Marseille, France}
\affil{$^{4}$INAF  -  Osservatorio  Astronomico  di  Bologna,  via  Ranzani  1,  I-40127, Bologna, Italy}
\affil{$^{5}$Max Planck Institut f\"ur Extraterrestrische Physik, Giessenbachstrasse 1, D-85748, Garching bei M\"unchen, Germany}
\affil{$^6$Institut d'Astrophysique de Paris, CNRS \& UPMC, UMR 7095, 98 bis Boulevard Arago, 75014, Paris, France}
\affil{$^{7}$Institute for Astronomy, Swiss Federal Institute of Technology, 8093 Z\"urich, Switzerland}
\affil{$^{8}$ Institute for Astronomy, 2680 Woodlawn Dr., University of Hawaii, Honolulu, HI 96822, USA}
\affil{$^{9}$Subaru Telescope, 650 N. A`ohoku Place, Hilo, HI 96720, USA}
\affil{$^{10}$Department of Physics and Astronomy, University of California, Riverside, CA 92521, USA}
\affil{$^{11}$Kavli Institute for the Physics and Mathematics of the Universe (WPI), The University of Tokyo Institutes for Advanced Study, The University of Tokyo, Kashiwa, Chiba 277-8583, Japan}


\affil{\textit{submitted to ApJ}}

\email{afaisst@ipac.caltech.edu}




\begin{abstract}
We measure a relation between the depth of four prominent rest-UV absorption complexes and metallicity for local galaxies and verify it up to $z\sim3$. We then apply this relation to a sample of $224$ galaxies at $3.5 < z < 6.0$ ($\left<z\right> = 4.8$) in COSMOS, for which unique UV spectra from DEIMOS and accurate stellar masses from SPLASH are available.
	The average galaxy population at $z\sim5$ and $\logm > 9$ is characterized by $0.3-0.4~{\rm dex}$ (in units of $\oabund$) lower metallicities than at $z\sim2$, but comparable to $z\sim3.5$. We find galaxies with weak/no \lya emission to have metallicities comparable to $z\sim2$ galaxies and therefore may represent an evolved sub-population of $z\sim5$ galaxies.
	We find a correlation between metallicity and dust in good agreement with local galaxies and an inverse trend between metallicity and star-formation rate (SFR) consistent with observations at $z\sim2$. The relation between stellar mass and metallicity (MZ relation) is similar to $z\sim3.5$, however, there are indications of it being slightly shallower, in particular for the young, \lya emitting galaxies. 
	We show that, within a ``bathtub'' approach, a shallower MZ relation is expected in the case of a fast (exponential) build-up of stellar mass with an $e-$folding time of $100-200\,{\rm Myr}$. Due to this fast evolution, the process of dust production and metal enrichment as a function of mass could be more stochastic in the first billion years of galaxy formation compared to later times.

\end{abstract}





\section{Introduction}

Metallicity is an important diagnostic for understanding the details of galaxy formation, because its connection to the history of a galaxy's star-formation rate (SFR), gas in- and outflows -- i.e., the interplay between the inter-stellar medium (ISM) and the inter-galactic medium (IGM). While the metal content of galaxies and its relation to other physical properties has been studied in depth at $z\sim2-3$, there are only a handful of observations at higher redshifts.

	The presence of a tight relation ($0.07-0.20~{\rm dex}$ scatter) between the galaxy's stellar mass and gas-phase metallicity (the MZ relation), measured by the ratio of oxygen to hydrogen\footnote{We define the solar metallicity as $\Zsol=0.02$.}, has been seen in local galaxies as early as in the 1970's \citep[][see also \citet{BOTHWELL13}]{LEQUEUX79}. Recently improved instrumental capabilities in the near-infrared (near-IR) allowing the measurement of metallicity at higher redshifts, show that this relation holds up to $z\sim2$ \citep{SAVAGLIO05,MAIER05,ERB06,MAIOLINO08,ROSEBOOM12,HENRY13,GALLAZZI14,WUYTS14,MAIER15,SANDERS15,SALIM15} and even up to $z\sim3$ although only by small sample sizes \citep[][Onodera et al. 2015, in prep]{MANNUCCI09,BELLI13,MAIER14}.
	At all redshifts, the metal content of galaxies is observed to increase with increasing stellar mass but flattens out above a stellar mass of roughly $\logm=10.0-10.5$. The MZ relation is also observed to evolve with redshift with galaxies at high redshifts showing a lower metal content. Galaxies at $z\sim3$ are found to have only $\sim$1/5$^{\rm th}$ of the solar metal abundance \citep[][Onodera et al. 2015, in prep]{MANNUCCI09,JONES12}. Besides the correlation of metallicity with stellar mass and cosmic time, an \textit{inverse} dependence with SFR has been found in local galaxies \citep{ELLISON08,LARALOPEZ10,ANDREWS13}. This has led to the so-called "fundamental mass-metallicity relation" \citep[see][]{MANNUCCI10}, however, the universality of this 3-dimensional relation is debated as there are evidences of it breaking down at $z\sim2-3$ \citep[see e.g.,][]{MAIER14,SALIM15}
	
	Several physical processes have been suggested for the origin of the MZ relation.
	For example, the deficit of metals in low mass galaxies can be attributed to strong winds (stellar winds or supernovae feedback) that excavate metal-rich gas out of the galaxy's low gravitational potential \citep[e.g.,][]{LARSON74,EDMUNDS90,GARNETT02,TREMONTI04,DELUCIA04,FINLATOR08}. Such strong outflows are found to be ubiquitous in local star-burst galaxies as well as star-forming galaxies at higher redshifts \citep[e.g.,][]{STEIDEL10,MARTIN12,KORNEI12}.
	Also, more massive galaxies tend to form their stars earlier \citep[an effect called "Downsizing", e.g.,][]{COWIE96,GAVAZZI96,FRANCESCHINI06,PEREZGONZALEZ08,ILBERT13} and
	therefore start to enrich their ISM earlier than less massive galaxies.
	It is also important to mention that the shape of the initial stellar mass function (IMF) affects the rate at which the ISM gets enriched by metals and thus has a direct influence on the shape of the MZ relation \citep[e.g.,][]{KOEPPEN07}.
	The evolution of metallicity with cosmic time and its dependence on stellar mass and SFR has been successfully predicted by semi-empirical models under the assumption that the metal content in the ISM is set by the balance between the inflow of pristine (i.e., metal poor) gas and enrichment through star formation. \citep[see][]{BOUCHE10,DAVE12,LILLY13,PIPINO14,HARWIT15,FELDMANN15}.
	 
	 The gas-phase metallicity of a galaxy is commonly derived from the ratio between strong emission line features in the optical part of spectrum (\oii, \oiii, \halpha, \hbeta, \nii) calibrated to theoretical models \citep[e.g.,][]{NAGAO06,KEWLEY02,KEWLEY08,MAIOLINO08}.
	The most common of these so-called ``strong-line methods" include the R$_{23}$ diagnostics \citep[combining \oiii, \oii, and \hbeta;][]{PAGEL79} as well as the N2 method \citep[combining \halpha~and \nii;][]{STORCHIBERGMANN94} that is commonly used to break the degeneracy and dust dependence of the $R_{23}$ method.
	While the strong-line methods can be employed up to $z\sim3$, at higher redshifts the diagnostic lines fall out of the wavelength window of ground based near-IR spectrographs.
	This considerably hampers the investigation of the metal content of galaxies at very early epochs and therefore our understanding of the formation of these galaxies, until the advent of the \textit{James Webb Space Telescope} (JWST).
	
	A correlation between metallicity and the equivalent width (EW) of absorption features in the rest-frame ultra-violet (UV) is expected from theoretical models \citep[e.g.,][]{ELDRIDGE12} and is observed in star-burst galaxies in the local universe \citep[e.g.,][]{HECKMAN98,LEITHERER11}. It provides an alternative way to probe statistically the metal content of galaxies. This method has already been used to determine stellar and gas-phase metallicities of $z\sim3$ galaxies returning reasonable results \citep[e.g.,][]{MEHLERT02,SAVAGLIO04,MARASTON09,SOMMARIVA12}.
	\textit{However, the this correlation has never been directly verified to hold at high redshifts}.
	
	In this paper, we aim
	\textit{(i)} to verify the EW vs. metallicity relation at high redshift and
	\textit{(ii)} use it to investigate the metal content of $224$ star-forming galaxies at $z\sim5$ with rest-frame UV spectra obtained by the Deep Imaging Multi-Object spectrograph \citep[DEIMOS,][]{FABER03}. The large sample size allows us to investigate the dependence of metallicity on stellar mass for the first time at these early epochs, giving us clues on how these galaxies are formed.
	
	The paper is organized as follows:
	We first verify the EW vs. metallicity relation of local galaxies on a sample of $z\sim2-3$ galaxies (\myrefsec{s:calibration}).
	We then present in \myrefsec{s:data} the sample of $z\sim5$ galaxies to which we will apply this relation to estimate their metallicities.
	In \myrefsec{s:composite} we present the $z\sim5$ composite spectrum and describe the measurement of EWs including the correction for various biases. The results of our analysis are presented in \myrefsec{sec:results}.
	Eventually, in \myrefsec{sec:discussion}, we discuss the metal content of $z\sim5$ galaxies and conclude and summarize the results of this paper in \myrefsec{s:summary}.
	
	We adopt a flat cosmology with $\Omega_{\Lambda,0} = 0.7$, $\Omega_{m,0}=0.3$, and $h=0.7$. Magnitudes are given in the AB system \citep{OKE83} and stellar masses are computed for a \citet{CHABRIER03} initial mass function. Metallicities are quoted in the calibration of \citet{MAIOLINO08}\footnote{Note that different calibrations can lead to up to $0.3~{\rm dex}$ different metallicity measurements for single galaxies \citep[e.g.,][]{KEWLEY08}.}.

\begin{figure*}[t!]
\centering
\includegraphics[width=1\columnwidth, angle=270]{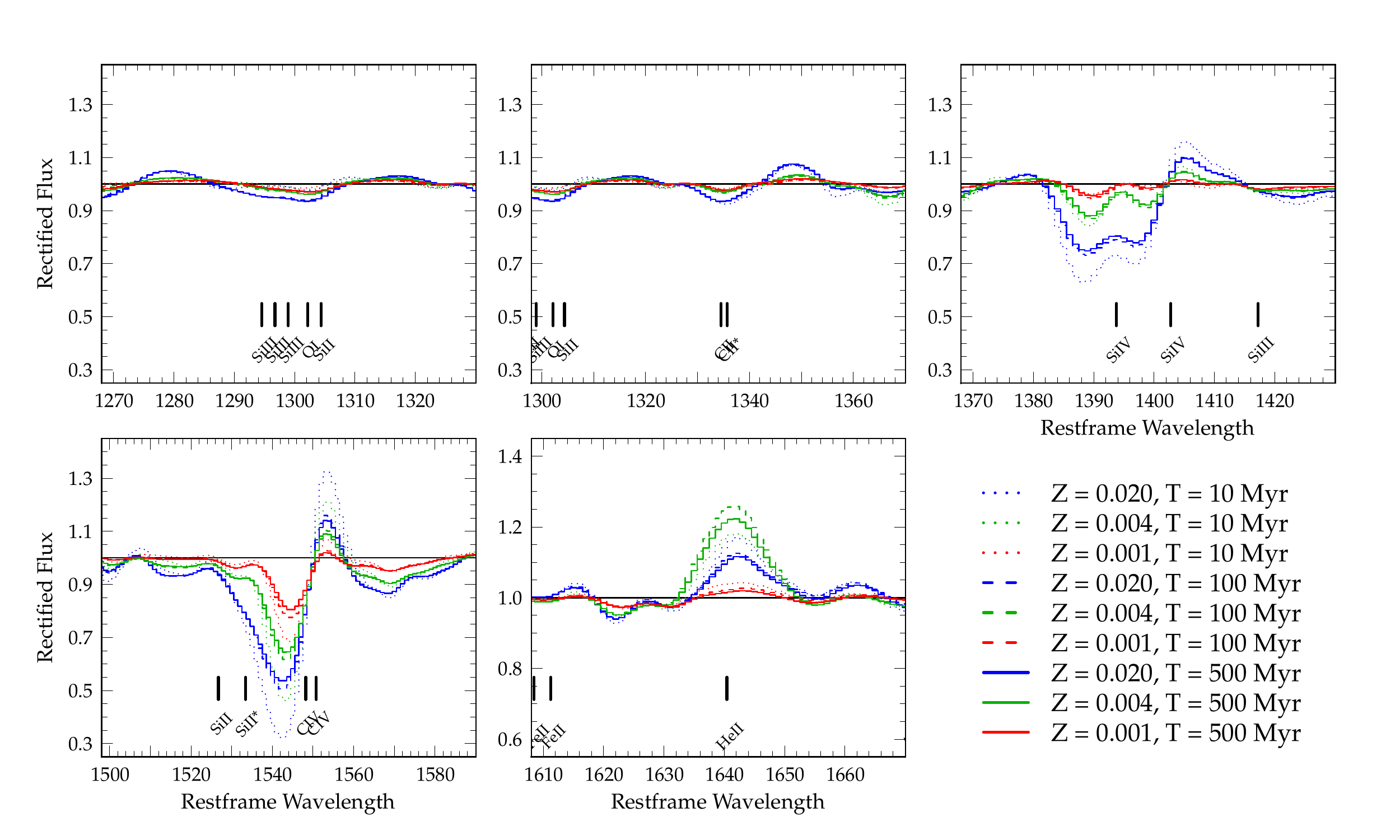}
\includegraphics[width=1\columnwidth, angle=270]{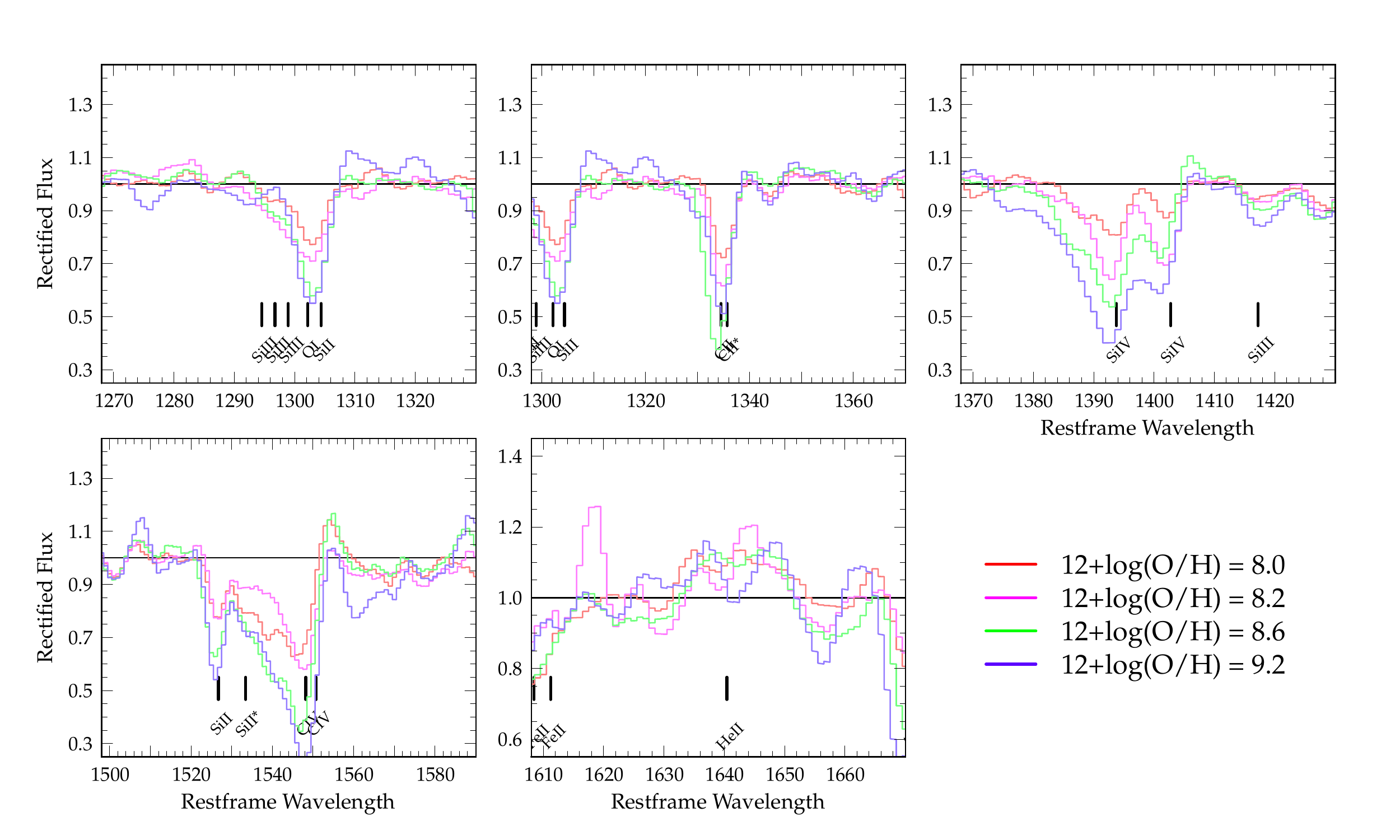}
\caption{\textbf{The top five panels} show absorption line features as well as \heii~emission as a function of metallicity and age for synthetic spectra generated by the \texttt{BPASS} code. Shown are models with metallicities $Z=0.02$ (blue), $Z=0.004$ (green), and $Z=0.001$ (red). These metallicity abundances are shown for a 10, 100, and 500~$\myr$ old stellar population, shown in dotted, dashed, and solid lines, respectively. Notice that above $\sim100~\myr$ there is no significant dependence of the absorption features on age.
\textbf{The bottom five panels} show the same wavelength ranges but of 22 observed spectra of local star-burst and star-forming galaxies by \citet{LEITHERER11}. The spectra are split and stacked in four different metallicity bins given their oxygen abundances measured by various authors \citep[see][]{LEITHERER11}. The general trends of absorption depth with metallicity are consistent with the \texttt{BPASS} models except for the ISM lines that are not included in this model.
\label{fig:absorptionmodel}}
\end{figure*}

\section{UV absorption lines as a measure of metallicity}\label{s:calibration}

	The strong-line methods, based on the ratio of strong optical emission lines to determine gas-phase metallicities work well up to redshift of $z\sim3$. Above that the lines are shifted out of the wavelength range of current ground-based near-IR spectrographs.
	An alternative way of estimating metallicity is to use the relation between the EW of inter-stellar (IS) and photospheric absorption features in the rest-frame UV.
	In this section, we calibrate this relation using a set of local galaxies and verify it at redshifts as high as $z\sim3$, before applying it to a sample of $z\sim5$ galaxies.

	There are several physical aspects that could cause a correlation between metallicity and EW of UV absorption lines. For details we refer to \citep[][]{HECKMAN98}, here we briefly summarize the main points.
	
	\begin{itemize}
	\item In general, the spectral properties are set by OB stellar populations. The evolutionary history of these correlations strongly depends on the stellar mass-loss rates, which in turn depend on the metal abundance \citep[][]{MAEDER94}.
	
	\item Winds from hot stars contribute to the line profiles of \civ~and \siiv. As shown theoretically (and also confirmed observationally), the wind strength of these stars is metal dependent \citep[see][]{CASTOR75,WALBORN95}.
	
	\item Dust extinction is proportional to metallicity and sets the column density for producing the absorption features \citep[e.g.,][]{HECKMAN98}.
	
	\item Most of the IS lines are optically thick at average columns densities of the ISM in star-forming galaxies \citep[e.g.,][]{HECKMAN97,SAHU97,PETTINI95}. Therefore, their line depths depend only weakly on the column density but more strongly on the velocity dispersion of the absorbing gas.
	
	\item The more metal rich star-bursts are, the more powerful in terms of their UV and IR luminosities. Moreover, they reside in more massive galaxies with higher average ISM velocity dispersion caused by both supernovae and stellar winds. The velocity dispersion widens the optically thick IS lines.

	\end{itemize}

	While the EW vs. metallicity relation is clearly seen in local galaxies, it has not been verified at higher redshifts. In the following, we combine the data from more than 50 local galaxies and use $\sim20$ galaxies at $z\sim2-3$ to test the relation at high-z.
	We focus on the strongest lines observed in the spectra of high-z galaxies, which are: the photospheric/IS \siiii~complex at $1300~{\rm \AA}$, the IS \cii~doublet at $1335~{\rm \AA}$ as well as the \siiv~and \civ~wind features.


\subsection{Synthetic stellar library \texttt{BPASS}}

We begin by investigating the relation between absorption line EW and metallicity from a theoretical perspective using a synthetic stellar library created by the Binary Population and Spectral Synthesis \citep[\texttt{BPASS},][]{ELDRIDGE09,ELDRIDGE12}\footnote{\url{http://www.bpass.org.uk}} models.
	These combine stellar evolution models with libraries of synthetic atmospheric spectra and to provide high-resolution modeling of stellar populations. They also include the binary evolution of stars, resulting in stellar populations that are bluer and older compared to populations of stars without binaries. In addition, the models are post-processed by \texttt{Cloudy} \citep{FERLAND98} to include nebular emission.  
	These models allow us to estimate the correlations of the EW with other physical quantities like the age of stellar population. In the following, we use the \texttt{BPASS} models with constant star-formation and the contribution of binary stars.
	
	We use 9 different model populations with different stellar population ages and metallicities. These consist of a old ($500~\myr$), intermediate age ($100~\myr$), and young ($10~\myr$) stellar populations with metallicities of $Z=0.001, 0.004,~{\rm and}~0.02$.	
	In the top five panels of \autoref{fig:absorptionmodel}, we show the regions around absorption line complexes and \heii~from \texttt{BPASS} generated models. The solid, dashed, and dotted lines correspond to old, intermediate age, and young stellar populations and the colors show different metallicities. The correlation between IS absorption and metallicity is clearly evident. Also, the models are largely independent of age for stellar populations older than $\sim100~{\rm Myrs}$.
	The \heii~emission variations with metallicity are more complicated so \heii~is not used in our analysis.
	
\begin{figure*}[t!]
\centering
\includegraphics[width=1.6\columnwidth, angle=0]{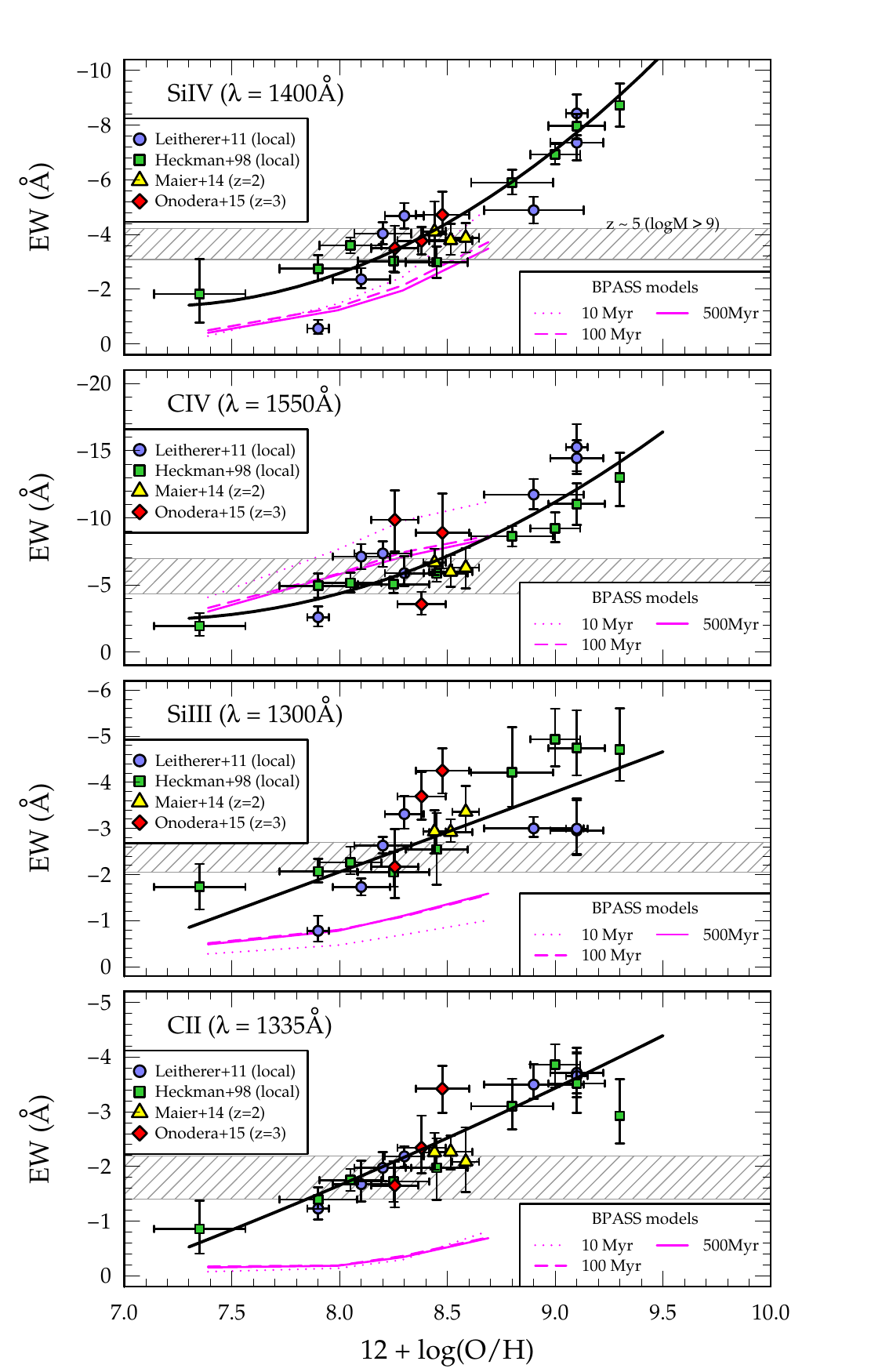}
\caption{Correlation between the rest-frame EW of UV absorption features at $1200-1600~{\rm \AA}$ and gas-phase metallicity.
The dotted, dashed, and solid magenta lines show \texttt{BPASS} models with stellar population ages of $10$, $100$, and $500$ Myrs, respectively.
The symbols show galaxy samples with observed rest-frame UV absorption features and metallicities measured by the strong-line methods. These samples have been compiled at different redshifts and are binned in metallicity: local \citep[blue squares, green circles;][]{HECKMAN98,LEITHERER11}, $z\sim2$ \citep[yellow triangles;][]{MAIER14}, and $z\sim3$ (red diamonds; Onodera et al. in prep.).
Note that the \texttt{BPASS} models under-estimated the EW of low-ionization IS absorption features that are mainly produced in the ISM of galaxies that are not included in the models (see text).
We verify that the locally observed EW vs. metallicity relation to hold up to redshift of $z=2-3$. The fitted parameters are given in \autoref{tab:fit}, the fit is shown as black solid line.
The hatched region shows the range of EWs for our $z\sim5$ galaxies with $\logm > 9.0$.
\label{fig:EW}}
\end{figure*}

	\autoref{fig:EW} shows the above mentioned trends more quantitatively by the rest-frame EW measured for different absorption features as a function of metallicity\footnote{We converted \texttt{BPASS} stellar metallicities into oxygen abundances assuming the standard conversion $\oabund = 8.69+\log(Z/0.02)$.}. The \texttt{BPASS} models are represented as magenta dotted, dashed, and solid lines for stellar populations of 10, 100, and 500~Myrs. As mentioned above, the change in EW with age for stellar populations older than $\sim100~\myr$ is negligible. However, for younger galaxies we expect an increased \civ~absorption and a deficit for the \siiii~complex. The \cii~absorption, on the other hand, remains unchanged with age even for very young stellar populations. In high S/N spectra, the ratio between \cii~and \civ~could therefore be used as an additional constraint on the age of the underlaying stellar population.

\subsection{Calibration on local galaxies}

We consider two partially overlapping samples of local star-burst galaxies with UV spectral coverage between $1000~{\rm \AA}$ and $2000~{\rm \AA}$ and measured metallicities from, if available, electron temperature (ratio of \oiii~line ratios at different wavelengths) and strong-line methods (using the \citet[][]{EDMUNDS84} calibration) to investigate the relation between IS absorption line EWs and metallicity.
Both samples are based on the International Ultraviolet Explorer (IUE) data archives \citep[see][]{KINNEY93} and are presented in detail in \citet[][]{HECKMAN98} and \citet[][]{LEITHERER11}. Furthermore, we make sure that these samples do not contain AGNs.

	The \citet{HECKMAN98} sample consists of a compilation of 45 local star-burst galaxies with low-resolution (${\rm R}~\sim 200-300$) UV spectral coverage of $1150-2000~{\rm \AA}$ as well as metallicity measurements from the literature.  The metallicity of the sample ranges from $\oabund = 7.5$ to $\oabund = 9.5$. 
	In their study, \citet{HECKMAN98} show the correlation for individual galaxies between metallicity and \textit{average} IS (\siii, \oi, \siiii, and \cii) as well as wind (\civ~and \siiv) absorption EWs. However, since the \siiv~and \civ~lines show very different EWs we re-extract the spectra from the NED archive\footnote{\url{http://ned.ipac.caltech.edu/}} and measure the EWs of the lines separately on stacked spectra of different metallicity bins to be consistent with our methods.
	
	The \citet{LEITHERER11} sample consists of a compilation of 46 spectra in sub-regions of 28 local star-burst and star-forming galaxies\footnote{\url{http://www.stsci.edu/science/starburst/templ.html}} with moderate-resolution (${\rm R}~\sim1300$) UV spectral coverage of $1150-3200~{\rm \AA}$ as well as metallicity measurements from the literature.
	The spectra were obtained by the Faint Object Spectrograph \citep[FOS,][]{HARMS79} as well as the Goddard High Resolution Spectrograph \citep[GHRS,][]{BRANDT94} onboard the Hubble Space Telescope (HST). The metallicity of the sample ranges from $\oabund=7.5$ to $\oabund=9.5$.
	Out of these 46 spectra, we have carefully chosen 22 for which there is good data coverage across the spectral features examined in the previous section and that show no emission due to the \oi~airglow line at $1304~{\rm \AA}$ as this would impact the measurement of the \siiii~absorption complex at $1300~{\rm \AA}$. The lower panels of \autoref{fig:absorptionmodel} show the stacked spectra in four different metallicity bins in the same five spectral ranges as for the \texttt{BPASS} models. The composites quantitatively show the connection between strong IS absorption and metallicity. Furthermore, the blue wing of \civ~is more pronounced for metal rich systems in agreement with stellar winds having higher terminal velocities in high metal abundance regions as expected from the \texttt{BPASS} models.
		
	\autoref{fig:EW} shows the correlation between EW and metallicity for the Leitherer et al. (blue circles) and Heckman et al. (green squares) galaxies binned in metallicity. The horizontal error bars represent the dispersion of metallicities in each bin and the vertical error bars are derived from a Monte-Carlo sampling accounting for the uncertainties in the measurements of the continuum.
	We find a relatively tight, monotonic relation between the EWs and the metallicity in good agreement with the \texttt{BPASS} models for the wind dominated lines \siiv~and \civ. The models, however, do not treat the ISM part of absorption and therefore under-estimate the EW of the low-ionization (pure IS) absorption complexes around \siiii~and \cii~(the same is partially true for \siiv).

\subsection{Galaxies at $z\sim2-3$}

To verify the local relation at higher redshifts, we repeat the previous test at $z\sim2-3$.
	First, we use a sample of $20$ galaxies at $2.1 < z < 2.5$ presented in \citet{MAIER14} with zCOSMOS-deep \citep[see][]{LILLY07} UV spectra in the observed wavelength range $3600-6800~{\rm \AA}$ 
	and metallicities estimated by a simultaneous fit (using the \citet[][]{KEWLEY02} models) to five optical emission lines (\oii, \hbeta, \oiii, \halpha, and \nii) accounting for dust, ionization parameter, and metallicity \citep[see][]{MAIER05}.
	
	For each of these galaxies, we retrieve fully calibrated zCOSMOS-deep UV spectra and measure the EW of the UV absorption features on composite spectra in bins of metallicity as it will be outlined in the later sections where we describe the measurements on our $z\sim5$ galaxy sample.
	The result is shown in \autoref{fig:EW} (yellow triangles). The measurements at $z\sim2$ agree well with the local EW vs. metallicity relation at $\oabund \sim 8.5$.


A further test is conducted using a sample of 11 galaxies at $z\sim3$ with metallicities ranging between $8.0 < \oabund < 8.8$ from Onodera et al. (in preparation). These galaxies were observed in near-IR using the Multi-Object Spectrometer for Infra-Red Exploration \citep[MOSFIRE;][]{MCLEAN12} and metallicities are computed with the \citet{MAIOLINO08} calibration using simultaneously strong-line methods including R$_{23}$ and line ratios \oii/\oiii, \oii/\hbeta, \oiii/\hbeta, and \neiii/\oii. The EWs of UV absorption complexes are derived from zCOSMOS-deep UV spectra using the same methods as for the $z\sim2$ galaxies.
	The galaxies (stacked in three metallicity bins by a running mean) are shown with red diamonds in \autoref{fig:EW}. They agree well with the local relation and the $z\sim2$ galaxies.
	
	Overall the EW vs. metallicity relation appears to hold up to $z\sim3$, a time range of 11 billion years.

\capstartfalse
\begin{deluxetable}{lccc}
\tabletypesize{\scriptsize}
\tablecaption{Fit to the EW vs. metallicity relation according to ${\rm EW} = p_{0} + p_{1}Z + p_{2}Z^{2}$.\label{tab:fit}}
\tablewidth{0pt}
\tablehead{
\colhead{Aborption} & \colhead{$p_{0}$} & \colhead{$p_{1}$} & \colhead{$p_{2}$} 
}
\startdata
\siiii~ ($1300~{\rm \AA}$) &	11.371$^{+1.918}_{-2.314}$ & -1.621$^{+0.503}_{-0.515}$ & -0.007$^{+0.029}_{-0.035}$  \\[0.2cm] 
\cii~ ($1335~{\rm \AA}$) &	5.479$^{+2.027}_{-2.126}$ & -0.102$^{+0.487}_{-0.506}$ & -0.098$^{+0.028}_{-0.031}$  \\[0.2cm] 
\siiv~ ($1400~{\rm \AA}$) &	-86.288$^{+4.209}_{-4.191}$ & 23.802$^{+0.879}_{-1.071}$ & -1.665$^{+0.057}_{-0.056}$  \\[0.2cm] 
\civ~ ($1550~{\rm \AA}$) &	-125.961$^{+5.584}_{-5.966}$ & 34.797$^{+1.304}_{-1.454}$ & -2.449$^{+0.078}_{-0.084}$
\enddata
\end{deluxetable}
\capstarttrue

\subsection{The UV EW vs. metallicity relation}

Since the relation is verified to be valid up to $z~\sim~3$, we fit this relation for each UV absorption complex by a $2^{{\rm nd}}$-order polynomial function of the metallicity $Z$ (expressed in $\oabund$).
The fit is valid between $7.3 < \oabund < 9.5$. The coefficients for each of the UV absorption complexes are given in \autoref{tab:fit} and the best fit is shown in \autoref{fig:EW} as solid black line. We estimate the errors on the fit using a Monte-Carlo simulation including the errors on the EW measurements and the widths of the metallicity bins.
We now apply this relation to a sample of $z\sim5$ galaxies, which we describe in the following section.

\section{Sample of $z\sim5$ galaxies}\label{s:data}

\subsection{DEIMOS campaign on the COSMOS field}

The data is based on the spectroscopic follow-up campaign of the Cosmic Evolution Survey \citep[COSMOS,][]{SCOVILLE07} using the Deep Imaging Multi-Object Spectrograph \citep[DEIMOS,][]{FABER03} on the Keck II telescope.
	This campaign was designed with the goal of building up a sample of high redshift galaxies that spans a wide range of stellar masses, dust, and galaxy activity, in order to study the mass assembly, black hole growth, and feedback processes that control star-formation in the early universe.
	In total, there are about 1,500 sources at redshifts $z>3.5$, including galaxies, AGNs, as well as radio sources.
	The diverse sample consists of galaxies selected by photometric redshifts, continuum-selected Lyman Break galaxies dropping out in the $B-$, $g-$, $V-$, $r-$, $i-$, and $z-$band broad-band filters, narrow-band selected Ly$\alpha$ emitting galaxies in IA624, NB711, and NB816, and galaxies selected in the infrared as well as radio\footnote{For the full set of filters available for COSMOS we refer to \citet{CAPAK07} and the official COSMOS web-page (\url{http://cosmos.ipac.caltech.edu/}).}. For more details on the sample selection we refer to \citet{MALLERY12}.
	
	The observations of the targets were carried out over the course of three years between January 2007 and February 2010. The DEIMOS set-up is a G830 grating blazed at $8640~{\rm \AA}$ (optimal wavelength range is $5800-9800~{\rm \AA}$) with an OG550 blocker and $1\arcsec$ wide slitlets resulting in a resolution of $3.3~{\rm \AA}$ (or $R\sim2600$, $\delta v = 130~{\rm km\,s}^{-1}$). This is sufficient to distinguish the \oii$\lambda 3727$ doublet structure and provides secure redshifts. The average integration time per target is $3.5~{\rm hours}$ in blocks of $30~{\rm minutes}$. The raw spectra were reduced and processed by the DEEP2 data reduction pipeline, which was modified to accommodate dithering.
	The relative flux calibration was performed by using calibration stars (HZ44, GD71, Feige110) that were observed in the same configurations as the science masks in each individual night and for the absolute flux calibration we used the existing multi-wavelength photometry available on COSMOS.

\begin{figure}[t!]
\centering
\includegraphics[width=1.05\columnwidth, angle=0]{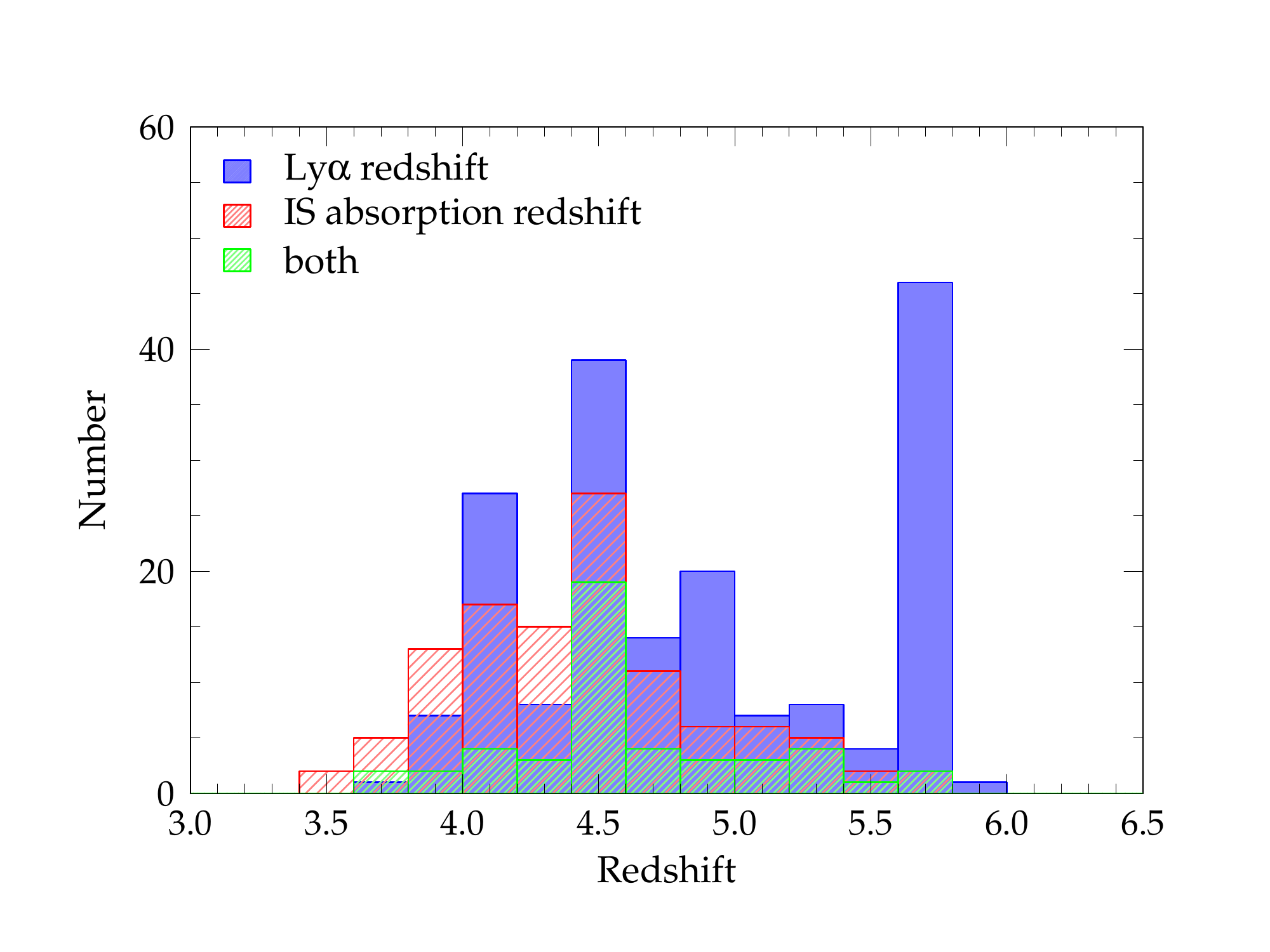}\\[-0.5cm]
\includegraphics[width=1.05\columnwidth, angle=0]{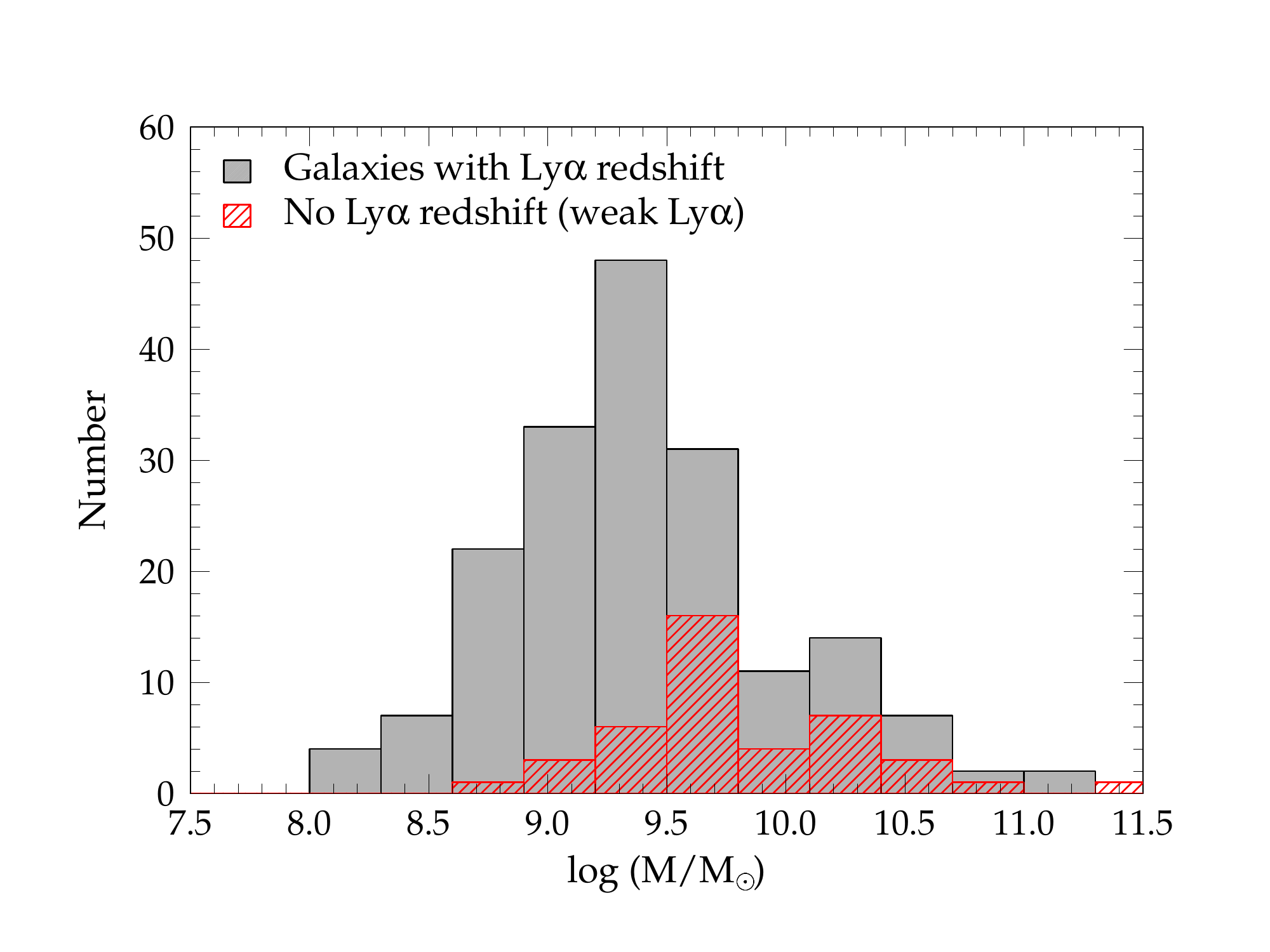}
\caption{\textbf{Top}: Redshift distribution for galaxies with \lya detection -- the final sample used this work (blue) measured by fitting a skewed gaussian to the emission line. The red hatched histogram shows the distribution of redshifts for single galaxies determined from IS absorption features (\siii, \cii, \siiii, \civ). The green hatched histogram shows galaxies for which redshifts from \lya and IS absorption features are available.
\textbf{Bottom}: The stellar mass distribution of galaxies with \lya redshifts (gray) and with only weak or no \lya emission (red). The stellar mass are obtained by SED fitting of 32-band COSMOS photometry including deep Spitzer/IRAC photometry \citep{LAIGLE15} to a comprehensive template library using the publicly available code \texttt{Le Phare}. The stellar masses are measured for a \citet{CHABRIER03} IMF.
\label{fig:samplestats}}
\end{figure}

\subsection{Sample selection and spectroscopic redshifts}

\subsubsection{General sample selection}

The spectroscopic redshifts of the galaxies are determined to 1st order by their \lya emission as well as low- and high-ionization absorption lines \citep[the spectra are analyzed manually by using \texttt{SpecPro};][]{MASTERS11}. The Lyman break detected by broad-band photometry is simultaneously used to support the redshifts. A quality flag between $1$ (very uncertain redshift) and $4$ (very certain redshift) is assigned based on the number of people (three in total) agreeing on the visually estimated redshift. The redshift distribution peaks at $\left< z \right> \sim 1.0$ and $\left< z \right> \sim 4.5$.
	For the purpose of this work, we choose galaxies with $3.5 < z < 6.0$ (with $\left< z\right> \sim 4.8$, see \autoref{fig:samplestats}) in order to have access to their rest-frame UV wavelength range ($1200-1800~{\rm \AA}$). In an additional step we manually remove galaxies with (broad) \civ~emission at $1550~{\rm \AA}$ that indicates a substantial contribution from active galactic nuclei \citep[AGN, e.g.,][]{ALLEN98}.

\subsubsection{Determination of redshift}

	The determination of \textit{systemic} redshifts is difficult at high redshifts due to strong outflows and resonant scattering of UV spectral features. 
	For our sample of galaxies, the systemic indicators are either blended and too faint to be observed in individual galaxies (photospheric lines \siiii, \ciii, \niv~and \fev), or fall out of the wavelength range of ground-based near-IR spectrographs (nebular lines \halpha, \hbeta~and \oiii).
	
	The most prominent spectral features of high-z star-forming galaxies are the \lya emission, IS absorption lines (\siii, \cii, \siiv, \civ), and the blended photospheric \siii/\ciii/\oi~line complex at $1300~{\rm \AA}$.
	The photospheric lines are blended and cannot be used to derive systemic redshifts.
	The \lya emission and IS absorption lines are affected by galactic winds and therefore shifted red- and blue-ward, respectively, by up to $1000~\kms$ \citep[e.g.,][]{PETTINI00,PETTINI02,STEIDEL10,LEITHERER11}.
	These different velocity shifts have to be taken into account when stacking the spectra.
		
	For this work, we aim for a galaxy sample that is as complete as possible in photometric and spectral properties. Thus we include galaxies for which the \lya line is detected as well as galaxies that have weak or no \lya emission or even \lya in absorption.
	For the former, the redshifts ($z_{\rm Ly\alpha}$) are obtained by the fitting of a skewed gaussian profile to the \lya emission line thereby accounting for the absorption on the blue side. For a detailed description of this procedure, we refer to \citet{MALLERY12}.
	The redshifts of the latter ($z_{\rm IS}$) are measured from their IS absorption lines (\siiv~as well as \civ~and \cii).
	The top panel of \autoref{fig:samplestats} shows the distribution of \lya redshifts (blue), the distribution of IS absorption redshifts (red) as well as the redshift distribution for galaxies for which a redshift from \lya and IS absorption features is available (green). The spike at $z\sim5.7$ contains galaxies selected in narrow-band imaging and therefore strong \lya emission. We have checked that these galaxies do not bias our sample.
	
	In the following stacking analysis, we will treat these samples (i.e., galaxies with and without \lya emission) separately but will eventually combine them for the measurement of metallicity. This allows us to  investigate possible selection effects that would occur if only focusing on one of the two populations.
	Also, we note that galaxies have in general a diversity of velocity offsets between IS absorption lines and \lya emission. This complicates the stacking analysis and also affects quantities measured on the composite spectrum. We will investigate and discuss this later in this paper.


\subsection{Stellar masses}

\subsubsection{The SED fitting procedure}

The stellar masses for our galaxies are derived from the fitting of COSMOS 32-band photometry from far-UV to IR (including-broad and narrow-bands in the optical) to a large library of spectral energy distribution (SED) templates at fixed spectroscopic redshifts using the publicly available code \texttt{Le Phare} \citep{ARNOUTS99,ILBERT06,LAIGLE15}.
	In the following, we briefly outline the procedure to obtain stellar masses for our galaxies. For more details on the SED fitting procedure, the imaging data reduction, and the extraction of the photometry we refer to \citet{LAIGLE15} as well as \citet{CAPAK07}, \citet{MCCRACKEN12} and \citet{ILBERT10,ILBERT13}.
		The library of synthetic templates is based on \citet{BRUZUALCHARLOT03} stellar population synthesis models, assuming a \citet{CHABRIER03} initial mass function (IMF)\footnote{The conversion of stellar mass to a \citet{SALPETER55} IMF goes roughly as $M_{\rm salp}~\sim~M_{\rm chab}~\times 1.77$}. Star-forming templates include common emission lines (\oii, \oiii, \halpha, \hbeta, and \lya), which fluxes we define to be proportional to the UV luminosity \citep{KENNICUTT98}. Accounting for emission lines in SED fitting is crucial at $z\gtrsim4$ to avoid a systematic over-estimation of stellar masses (up to factors of two at $z\sim5-6$) as well as biases in the measurement of colors \citep[e.g.,][]{SCHAERER09,STARK13,WILKINS13,GONZALEZ14,BARROS14,SANTINI15,FAISST15}.
	We use exponentially declining star-formation histories (SFHs) with $\tau=[0.3,1,3,5,30]~{\rm Gyr}$ along with delayed models (${\rm SFR}(t)\propto \tau^{-2} t e^{-t/\tau}$) in which the peak of star formation happens after $\tau=[0.1,0.5,1,3]~{\rm Gyr}$. We take into account different metallicities ($\Zsol$ and $0.2\Zsol$), and vary ${\rm E(B-V)}$ between $0$ and $0.8$. For the extinction curve we assume the following parameterizations: a $\lambda^{0.9}$ law \citep[see][]{ARNOUTS13} and a \citet{CALZETTI00} parametrization including a distinct $2175~{\rm \AA}$ feature that is observed in the extinction curve of the Milky Way and Large Magellanic Cloud as well as in galaxies at $z=2-4$ \citep[e.g.,][]{SAVAGE79,FITZPATRICK89,SCOVILLE15}. We do not find a significant difference in the stellar mass estimates using the original \citet{CALZETTI00} dust extinction curve.
	The stellar masses (including active stars and remnants) are defined to be the median of the probability distribution function after marginalizing over the templates.
	The bottom panel of \autoref{fig:samplestats} shows the stellar mass distribution for galaxies with \lya emission (gray) and with weak or no \lya emission (red).
	
	The final sample that is used in the following consists of 224 galaxies in the redshift range $3.5 < z < 6.0$ and stellar masses $\logm > 8.0$. Of these, 182 are detected in \lya emission and 42 show weak or no \lya emission.

\begin{figure}[t!]
\centering
\includegraphics[width=1.4\columnwidth, angle=270]{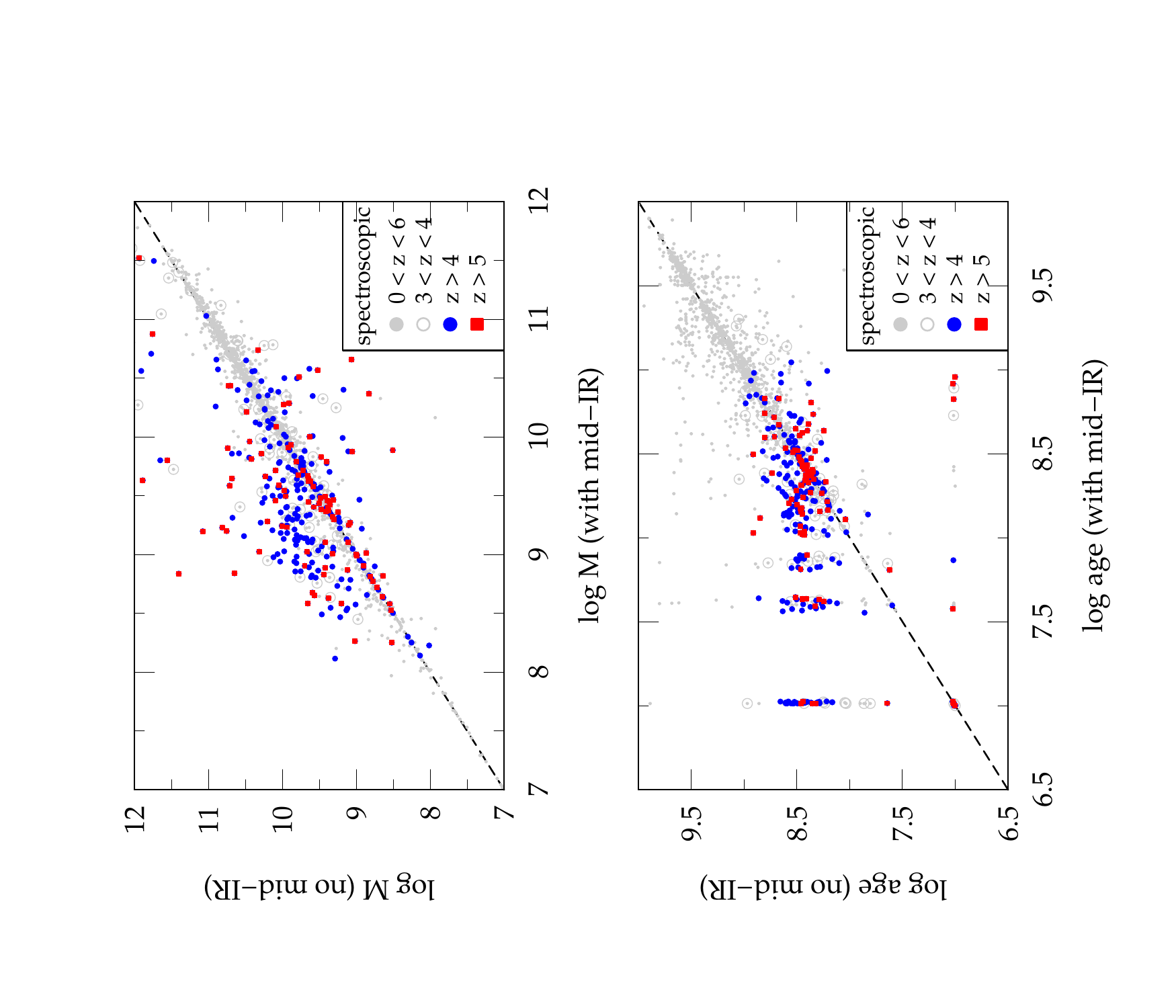}
\caption{The importance of deep mid-IR data for the computation stellar masses of $z>3.5$ galaxies. The two panels show the stellar masses and light-weighted ages of the stellar population from SED fitting including and omitting mid-IR data from Spitzer/IRAC. The symbols are colored as a function of spectroscopic redshifts (gray dots: $0 < z < 6$; gray open circles: $3 < z < 4$; blue points: $z > 4$; red squares: $z>5$).
Not using mid-IR data at $z>3$ results in over-estimated stellar masses by up to one magnitude at $\logm<10.0$ as well as over-estimated ages for young galaxies by more than one order of magnitude at $z\gtrsim3$.
\label{fig:IRAC}}
\end{figure}

\subsubsection{Aside: Importance of mid-IR photometry at $z>4$}

	At redshifts of $z\gtrsim3.5$, the observed optical to near-IR photometry does not cover the $4000~{\rm \AA}$ Balmer break, which is a sensitive measure of various parameters describing the stellar populations of a galaxy. Therefore, stellar mass estimates at high redshifts that lack mid-IR photometry can be severely biased.
	To overcome this problem, we use the \textit{Spitzer Large Area Survey with Hyper-Suprime-Cam} \citep[SPLASH, see][]{STEINHARDT14} on COSMOS. The SPLASH provides Spitzer/IRAC data at $3.6~{\rm \mu m}$ and $4.5~{\rm \mu m}$ that are $\sim2$ magnitudes deeper ($>25.5\,{\rm mag}$ at $3 \sigma$ in $3\arcsec$ apertures) than existing mid-IR data on COSMOS and therefore allows to measure reliable stellar masses at $z>4$ by providing a deep coverage of wavelength redder than the $4000~{\rm \AA}$ break. The Spitzer photometry is extracted using an improved version of IRACLEAN \citep{HSIEH12} in order to overcome the sources confusion (blending) in the Spitzer imaging data. For optimal de-blending, the combined $zYJHK_{{\rm s}}$ images have been used as a prior for the position and shape of the sources  \citep[for more details, see][]{LAIGLE15}.

		To demonstrate the importance of the mid-IR data, we have performed the stellar mass fitting with and without the addition of mid-IR data from SPLASH. 
		As shown in \autoref{fig:IRAC}, excluding mid-IR photometry results in an over-estimation of the stellar masses at $\logm<10.0$ by up to an order of magnitude at $z>3$. Stellar population ages are over-estimated by the same amount.
		This illustrates that without constraints in the mid-IR, the SED fitting of these high-z galaxies is dominated by the rest-frame UV part of the spectral energy distribution, i.e., directly proportional to the UV continuum slope $\beta$ with no constraints on the $4000~{\rm \AA}$ break.

\begin{figure*}[t!]
\centering
\includegraphics[width=2.1\columnwidth, angle=0]{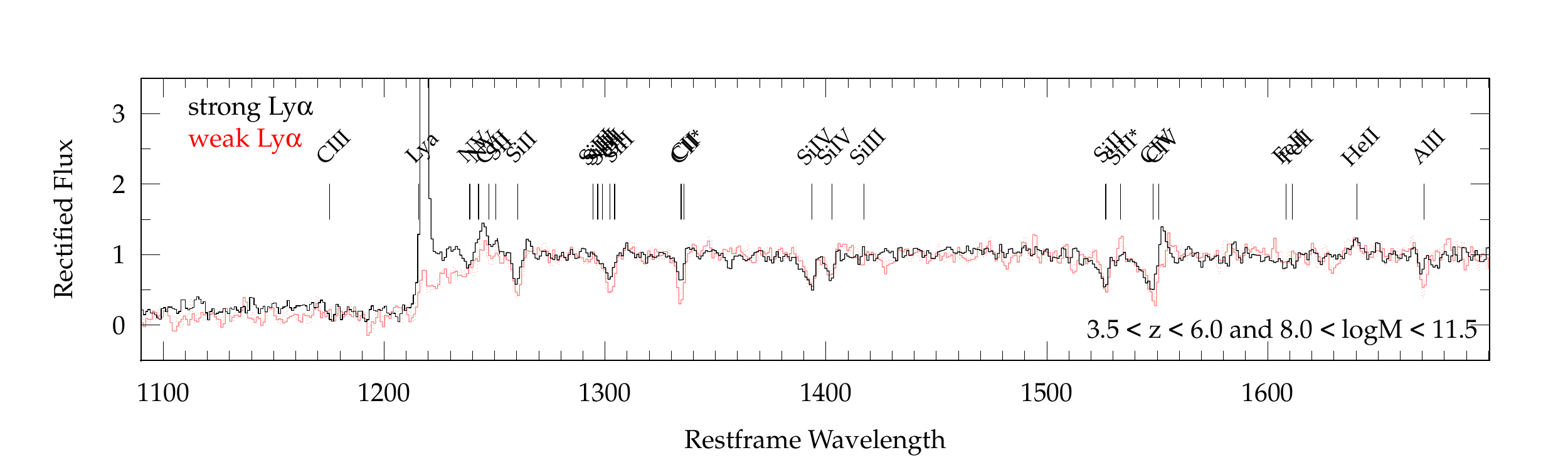}
\includegraphics[width=2.1\columnwidth, angle=0]{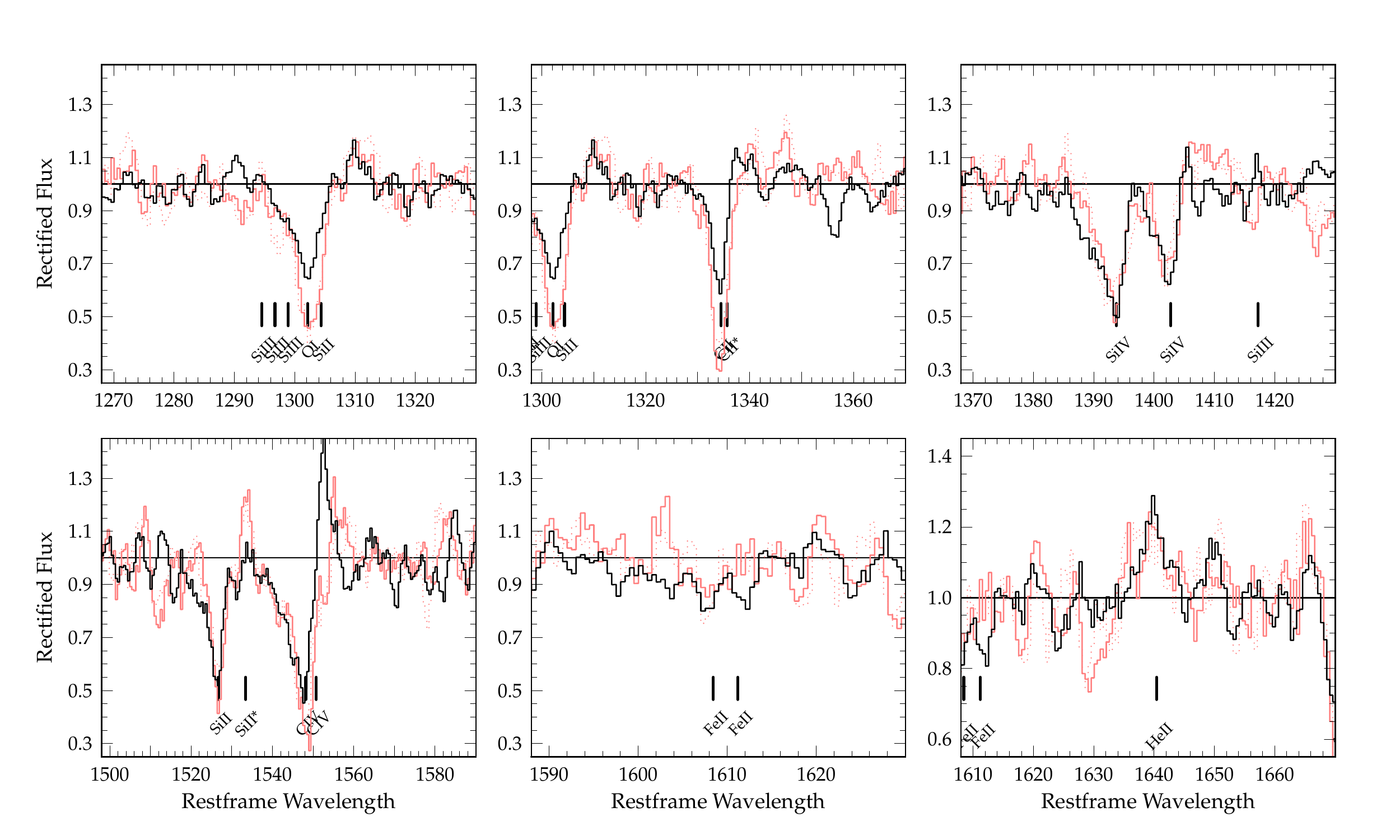}
\caption{Continuum-normalized composite spectrum for all galaxies at $3.5 < z < 6.0$ ($\left<z\right> = 4.8$) and $8.0 < \logm < 11.5$ (top) and zoom-in on the most prominent absorption complexes as well as \heii~(bottom). The black lines show galaxies with \lya emission. The solid red lines show the composite spectrum of galaxies with weak or no \lya emission, indicating their deeper absorption. The dashed red lines show the latter composite blurred with the average velocity dispersion $\dvlyais=429\pm230~{\rm km\,s^{-1}}$. The vertical lines name prominent absorption and emission features, the horizontal line on the zoom-in shows the continuum.
\label{fig:stackall}}
\end{figure*}


\section{Composite spectrum of $z\sim5$ galaxies}\label{s:composite}

In practice, investigations of spectral features as a function of physical parameters (e.g., stellar mass, see below) requires us to stack individual galaxy spectra because of the low S/N in individual spectra.
In this section we describe the stacking procedure and the measurements as well as bias correction of the EWs.

\subsection{Stacking procedure}

	Before stacking, we normalize each individual spectrum to the median flux measured between rest-frame $1250~{\rm \AA}$ and $1800~{\rm \AA}$. Within this wavelength range, we exclude regions of strong absorption lines, including the atmospheric A- and B-band (observed $7600-7630{\rm \AA}$ and $6860-6890{\rm \AA}$), the water absorption band (observed $>9000{\rm \AA}$), and the absorption lines \siii~($1255-1264{\rm \AA}$), \oi/\siii~($1290-1307{\rm \AA}$), \cii~($1326-1340{\rm \AA}$), \siiv~($1379-1405{\rm \AA}$), \siii~($1521-1529{\rm \AA}$), \civ~($1529-1553{\rm \AA}$), \feii~($1600-1613{\rm \AA}$), and \alii~($1663-1679{\rm \AA}$). We use the same wavelength windows to fit the continuum, which we use to rectify the individual galaxy spectra before stacking to account for their different UV continuum slope. The composite spectrum is created by the median stacking of the individual, rectified and normalized spectra in the rest-frame of the \lya emission line in the case of galaxies detected in \lya and in the rest-frame of IS absorption lines in the case of weak or no \lya emission.
	In the former sample (with \lya emission), the brightest and highest S/N galaxies with strong observed absorption features, however few in numbers, could bias the derived composite spectrum. We have checked, by removing 12 galaxies fulfilling this criteria, that the inclusion of these does not affect the results of this paper. 
	
	The binned and rectified $\left<z\right>=4.8$ composite spectrum in the rest-frame of \lya emission is shown in \autoref{fig:stackall} for galaxies with \lya emission (black line) and with weak or no \lya emission (red solid line). For a better comparison of the two stacks, we have shifted the latter into \lya rest-frame according to the average velocity offset between \lya and IS absorption lines measured for individual galaxies (see \autoref{app:velocity}). The most prominent spectral features are indicated by vertical lines.
	
	One could argue that the comparison of these two composites is unfair since the stacking in the rest-frame of the IS absorption lines may produces more enhanced absorption features than in the case of the composite in the rest-frame of \lya emission. This because galaxies show a distribution in velocity offsets between \lya and IS absorption lines that could blur the spectral features in the case of \lya stacks. To investigate this effect, we artificially blur the composite spectrum for galaxies with weak or no \lya emission by shifting the individual spectra randomly according to the gaussian velocity offset distribution with mean $\dvlyais=429\pm230~{\rm km\,s^{-1}}$ as derived in \autoref{app:velocity}. We find that this effect is negligible for the following results. The red dotted line shows one typical realization of this test as an example.
	The lower panels show a zoom-in on some of the spectral features, including the (blended) \siiii/\ciii/\oi~complex at $1300~{\rm \AA}$, the \cii~doublet at $1335~{\rm \AA}$, the \siiv~complex at $1400~{\rm \AA}$, the \siii/\civ~complex at $1540~{\rm \AA}$, the \feii~doublet at $1610~{\rm \AA}$, as well as the \heii~emission at $1640~{\rm \AA}$.

\capstartfalse
\begin{deluxetable}{cc}
\tabletypesize{\scriptsize}
\tablecaption{Dispersion in EW measurements due to variations in velocity offsets between \lya and IS absorption features estimated from 100 Monte-Carlo runs.\label{tab:EWdispMC}}
\tablewidth{0pt}
\tablehead{
\colhead{Absorption feature} & \multicolumn{1}{c}{Dispersion in EW ($1\sigma$, in ${\rm \AA}$)}\\
\colhead{~} & \multicolumn{1}{c}{for velocity-offset dispersion of $230\kms$}
}
\startdata
\siiv & 0.32 \\
\civ & 0.53 \\
\siiii & 0.18 \\
\cii & 0.10
\enddata
\end{deluxetable}
\capstarttrue


\subsection{Equivalent-widths}

\subsubsection{Measurement}

The rest-frame equivalent width (EW) of absorption features characterizes the physical properties of galaxies.
In this section, we measure EWs of different spectral features that we will later use to quantify the metal content of our galaxies.

The EW of a spectral features is defined as
	
\begin{equation}
{\rm EW} \equiv \int_\lambda \left(1 - \frac{f_{\lambda,{\rm feat}}}{f_{\lambda,{\rm cont}}}\right) d\lambda,
\end{equation}

\noindent
where $f_{\lambda,{\rm feat}}$ is the flux density across the absorption/emission feature and $f_{\lambda,{\rm cont}}$ is the flux density of the continuum (both in \ergscmA).
	Here, we define a "pseudo-continuum" across the spectral feature by a linear interpolation of the median continuum measured red- and blue-ward in a $10-20~{\rm \AA}$ window.
	In the case of absorption lines, we integrate over the pixels below this pseudo continuum and vice versa for emission lines.
	We find that the measurement of the pseudo continuum is the major source of uncertainty. This is especially true for galaxies at high redshifts and low S/N, and we therefore estimate the uncertainties of the measured EW by Monte-Carlo'ing over different pseudo continua, which are obtained including the errors in their flux densities.

\capstartfalse
\begin{deluxetable*}{lccc  cccc}
\tabletypesize{\scriptsize}
\tablecaption{List of prominent absorption features. See also \citet{LEITHERER11} and references therein.\label{tab:absfeat}}
\tablewidth{0pt}
\tablehead{
\multicolumn{8}{c}{ }\\[-0.1cm]
\multicolumn{4}{c}{--------------------- Line properties ---------------------} & \multicolumn{4}{c}{------------------ Measured on composite $z\sim 5$ spectrum$^{*}$ ------------------} \\[0.03cm]
\colhead{Feature} & \colhead{$\lambda_{{\rm vac}}$ [\AA]} & \colhead{$E_{{\rm ion}}$ [eV]} & \colhead{environment$^{a}$} & \colhead{$\lambda_{Ly\alpha}$ [\AA]$^{b}$} & \colhead{$\Delta v_{{\rm Ly\alpha-IS}}$$^{c}$} &\colhead{$\Delta v_{{\rm sys}}$$^{d}$} & \colhead{EW$_{{\rm rest}}$ [\AA]$^e$} \\[-0.2cm]
}
\startdata

\\[0.1cm]
\multicolumn{8}{c}{\textbf{Emission lines}}\\[0.1cm]
\hline\\[0.1cm]

\lya & 1215.67 & -- & IS & 1216.0 & 0 & 300 (340) $\pm170$ & -- \\
\heii & 1640.00 & -- & neb & 1638.0~$\pm~0.5$ & -370 & $-$70 (20) $\pm230$ & 1.7$^{+1.1}_{-0.7}$ \\[0.2cm]

\hline\\[0.01cm]
\multicolumn{8}{c}{\textbf{\siiii~absorption complex: 1290\AA -- 1310\AA}}\\[0.1cm]
\hline\\[0.1cm]
\siiii & 1294.54 & 16.35 & photo  & --$^{\dagger}$ & --$^{\dagger}$ & --$^{\dagger}$ & --$^{\dagger}$\\
\ciii & 1296.33 & 24.38 & photo  & --$^{\dagger}$ & --$^{\dagger}$ & --$^{\dagger}$ & --$^{\dagger}$ \\
\siiii & 1296.74 & 16.35 & photo  & --$^{\dagger}$ & --$^{\dagger}$ & --$^{\dagger}$ & --$^{\dagger}$ \\
\siiii & 1298.93 & 16.35 & photo  & --$^{\dagger}$ & --$^{\dagger}$ & --$^{\dagger}$ & --$^{\dagger}$ \\
\oi & 1302.17 & 0.00 & IS  & 1301.0~$\pm~0.5$ & $-$300  & 0 & --$^{\dagger}$ \\
\siii & 1304.37 & 8.15 & IS  & --$^{\dagger}$ & --$^{\dagger}$ & --$^{\dagger}$ & --$^{\dagger}$ \\[0.2cm]
\multicolumn{5}{c}{~} & \multicolumn{2}{r}{for this complex} & -2.2$^{+0.4}_{-0.4}$ \\[0.2cm]

\hline \\[0.01cm]
\multicolumn{8}{c}{\textbf{\cii~absorption complex: 1330\AA -- 1340\AA}}\\[0.1cm]
\hline\\[0.1cm]
\cii & 1334.53 & 11.26 & IS  & 1332.0~$\pm~0.5$ & $-$570 & $-$270 ($-$180) $\pm230$ & --$^{\dagger}$ \\
\cii$^{*}$ & 1335.71 & 11.26 & IS  & 1333.0~$\pm~0.5$ & $-$610 & $-$310 ($-$220) $\pm230$ & --$^{\dagger}$ \\[0.2cm]
\multicolumn{5}{c}{~} & \multicolumn{2}{r}{for this complex} & -1.6$^{+0.3}_{-0.3}$ \\[0.2cm]

\hline \\[0.01cm]
\multicolumn{8}{c}{\textbf{\siiv~absorption complex: 1385\AA -- 1410\AA}}\\[0.1cm]
\hline\\[0.1cm]
\siiv & 1393.76 & 33.49 & IS, wind  & 1391.0~$\pm~0.5$ & $-$590 & $-$290 ($-$200) $\pm230$ & -- \\
\siiv & 1402.77 & 33.49 & IS, wind  & 1401.0~$\pm~0.5$ & $-$590 & $-$290 ($-$200) $\pm230$ & -- \\[0.2cm]
\multicolumn{5}{c}{~} & \multicolumn{2}{r}{for this complex} & -4.1$^{+0.4}_{-0.4}$ \\[0.2cm]

\hline \\[0.01cm]
\multicolumn{8}{c}{\textbf{\civ~absorption complex: 1520\AA -- 1555\AA}}\\[0.1cm]
\hline\\[0.1cm]
\siii & 1526.71 & 8.15 & IS  & 1524.0~$\pm~0.5$ & $-$530 & $-$230 ($-$140) $\pm230$ & --$^{\dagger}$ \\
\feiv & 1530.04 & 30.65 & photo  & 1529.0~$\pm~0.5$ & $-$200 & 100 (190) $\pm230$ & --$^{\dagger}$ \\
\siii$^{*}$ & 1533.43 & 8.15 & IS  & -- & -- & -- & -- \\
\civ & 1548.19 & 47.89 & IS, wind  & 1543.0~$\pm~0.5$ & $-$1010 & $-$710 ($-$620) $\pm230$ & --$^{\dagger}$ \\
\civ & 1550.77 & 47.89 & IS, wind  & 1546.0~$\pm~0.5$ & $-$920 & $-$620 ($-$530) $\pm230$ & --$^{\dagger}$ \\[0.2cm]
\multicolumn{5}{c}{~} & \multicolumn{2}{r}{for this complex} & -6.4$^{+1.3}_{-1.4}$ \\[0.2cm]

\hline \\[0.01cm]
\multicolumn{8}{c}{\textbf{\feii~absorption complex: 1600\AA -- 1620\AA}}\\[0.1cm]
\hline\\[0.1cm]
\feii & 1608.45 & 7.87 & IS  & 1606~$\pm~0.5$ & $-$460 & $-$160 ($-$70) $\pm230$ & --$^{\dagger}$ \\
\feii & 1611.20 & 7.97 & IS  & 1610~$\pm~0.5$ & $-$230 & 70 (160) $\pm230$ & --$^{\dagger}$\\
\enddata
\tablenotetext{a}{Photospheric (photo), inter-stellar (IS), in winds of massive stars (wind), nebular (neb).}
\tablenotetext{b}{On the stacked spectrum in the rest-frame of \lya.}
\tablenotetext{c}{With respect to \lya in $\kms$. The errors are on the order of $\pm 100~\kms$.}
\tablenotetext{d}{With respect to systemic calibrated by \oi~in $\kms$. Values corrected for biases introduced by stacking (see text) are given in brackets.}
\tablenotetext{e}{Bias corrected, see text.}
\tablenotetext{$\dagger$}{Line is blended and no reliable wavelength, velocity offset or EW can be measured in this case.}
\tablenotetext{$^{*}$}{Galaxies with $3.5 < z < 6.0$ and $\logm>8.0$.}
\end{deluxetable*}
\capstarttrue


\subsubsection{Uncertainties, biases, and limitations}\label{sec:biases}

	There are several biases and uncertainties in the measurements of the EWs. First, individual galaxies show different velocity offsets of absorption lines to systemic. Stacking them in the rest-frame of the \lya emission line may cause a broadening of these absorption features. Second, there are uncertainties in the measurement of low EWs introduced by background noise. The latter does also set a detection limit. 

	We first investigate the effect of velocity offsets on the measured EW. For this we create 100 representations of a composite spectrum by shifting each individual spectrum according to the probability distribution of the velocity offsets derived in \autoref{app:velocity} and listed in \autoref{tab:absfeat}. For each representation, we measure the EW of the absorption features and compute the $1\sigma$ dispersion (see \autoref{tab:EWdispMC}). We find the corresponding error on the EW measurements to be less than $20\%$. This is smaller or similar to the measurement uncertainties of the EWs and therefore not a dominant source of error.
	
	Next, we investigate possible biases in the measurement of the EWs due to the effect of background noise. We create artificial composite spectra with gaussian absorption features (placed at $1330~{\rm \AA}$) for a grid of input EWs and include a noise level that we measure on the real composite spectra. We systematically under-estimate absorption EWs stronger than $1.5~{\rm \AA}$ by $\sim5\%$. EWs weaker than $1~{\rm \AA}$ are still detected but with a more substantial over-estimation of up to $\sim30\%$ on average. Although these biases are small compared to the uncertainties in the measurements, we correct for these in the following.
	
	In \autoref{tab:absfeat} we list the (corrected) rest-frame EW measured on the composite spectrum of $\logm > 8$ galaxies for five different absorption complexes (\siiii~($1300~{\rm \AA}$), \cii~($1335~{\rm \AA}$), \siiv~($1400~{\rm \AA}$), \civ~($1550~{\rm \AA}$), and \feii~($1610~{\rm \AA}$)) as well as \heii. Note, that the uncertainty on the latter (only observable in the highest redshift galaxies) is large, because of the decreasing number of galaxies and S/N of the spectra at high redshifts.


\section{Results}\label{sec:results}

\subsection{Dependences on EW}

Relations between metallicity (i.e., UV absorption EW) and dust, SFR, and stellar mass are expected from galaxies at lower redshifts ($z<2$). In the following, we focus the dependence of UV absorption EWs on SFR and dust attenuation.

We derive the line of sight dust attenuation for our galaxies from the UV continuum slope $\beta$ ($\log(f_{{\rm \lambda}}) \sim \beta \log(\lambda)$), which is fit on the COSMOS broad-band photometry between rest-frame $1300~{\rm \AA}$ and $2200~{\rm \AA}$. We restrict our sample to galaxies for which at least four filters are available for the fitting of $\beta$. The SFRs are derived from SED fitting and do not include (rest-frame) far-IR data, hence should therefore be taken with caution.

\begin{figure}[t!]
\centering
\includegraphics[width=1\columnwidth, angle=0]{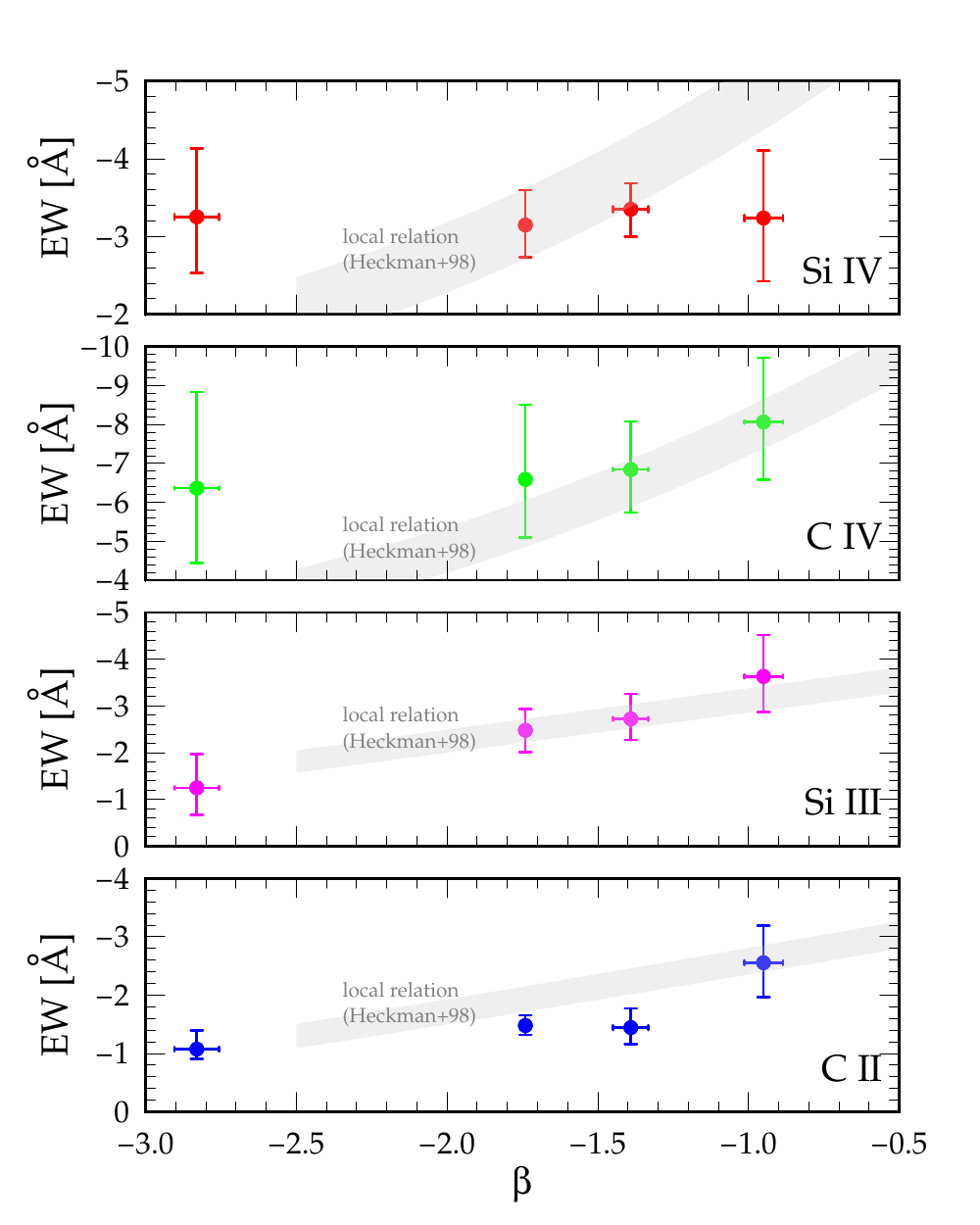}
\caption{Dependence of EW of UV absorption complexes on UV continuum slope $\beta$ (dust attenuation) measured from COSMOS broad-band photometry between $1300~{\rm \AA}$ and $2200~{\rm \AA}$. The points show our measurements at $z\sim5$, the gray bands show the relation measured for local galaxies \citep{HECKMAN98}. Both data agree well within $1\sigma$ for $\beta > -2$ (except for \siiv), indicating a correlation between dust and metal content in $z\sim5$ galaxies as it is seen in local galaxies.
\siiv~and \civ~are strongly affected by winds, which might reduce their correlation with dust attenuation.
\label{fig:EWbeta}}
\end{figure}

\begin{figure}[t!]
\centering
\includegraphics[width=1\columnwidth, angle=0]{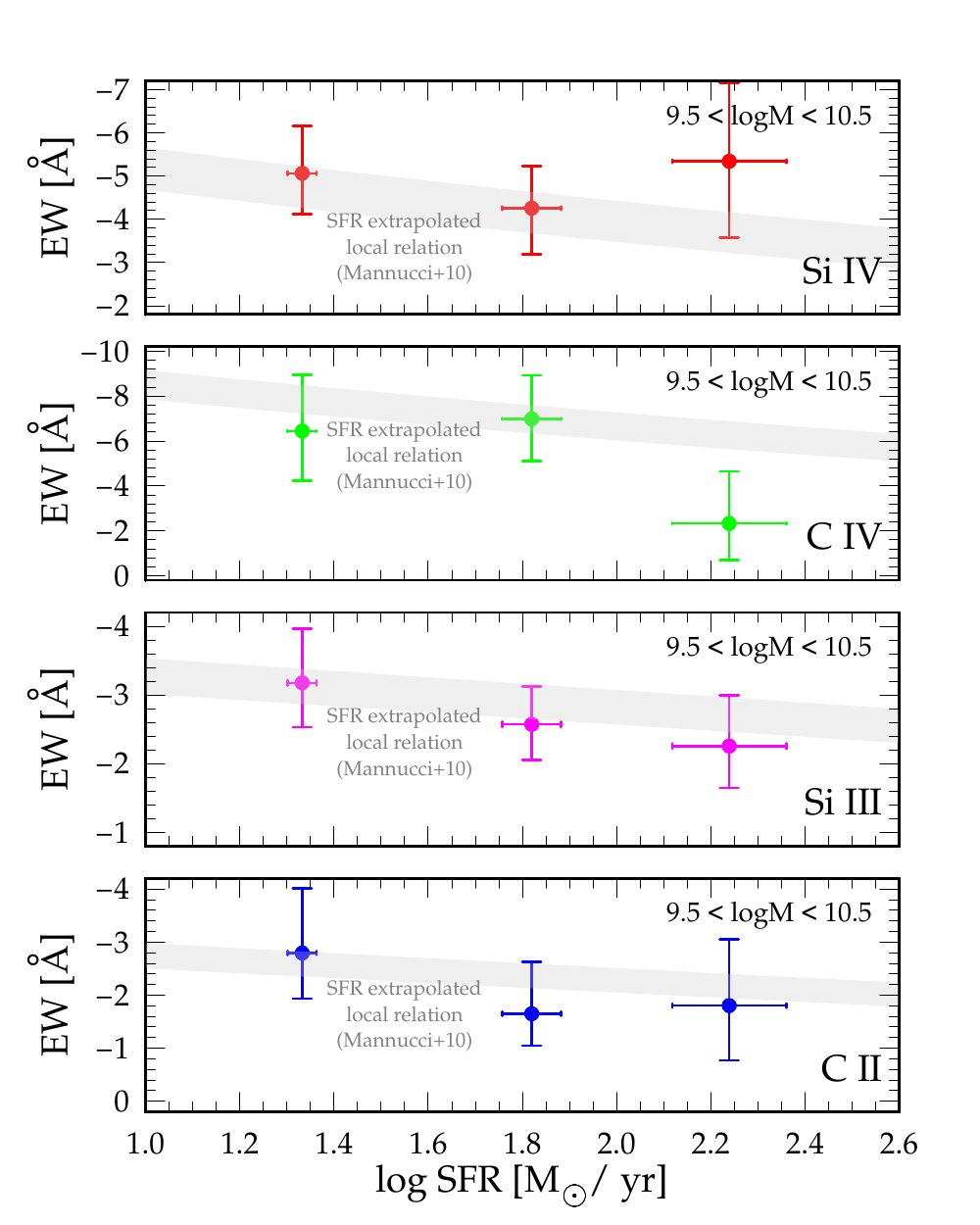}
\caption{Inverse correlation between EW of UV absorption complexes and star-formation (measured from SED fitting). The points show our measurements at $z\sim5$, the gray bands show the metallicity vs. star-formation relation for $z=0-2.2$ galaxies based on emission lines \citep{MANNUCCI10}. We have converted the metallicity to UV EW by using our relations presented in \autoref{tab:fit}. The data agree well within $1\sigma$, indicative that this inverse correlation at $z\sim5$ is real. This is expected in a picture where star-formation is help up by inflow of pristine (i.e., metal-poor) gas.
\label{fig:EWsfr}}
\end{figure}

\autoref{fig:EWbeta} shows the positive correlation between EW and dust for all absorption complexes except \siiv. This might be due to the strong wind component in this line. The gray band shows the EW vs. $\beta$ relation from local galaxies \citep{HECKMAN98} derived by inverting our EW vs. metallicity relation (see \myrefsec{s:calibration}). This comparison shows the very similar behavior of dust vs. metallicity between local and high-z galaxies.
\autoref{fig:EWsfr} shows the dependence of EW on SFR. We find a negative correlation, i.e., galaxies with high star-formation have a systematically lower EWs (or, equivalently, metallicities). This is expected in a picture where SFR is held up and fueled by the inflow of pristine (i.e., metal poor) gas. Also, metals can be expelled from (low mass) galaxies by strong outflows due to their high star-formation. The gray band shows the expected dependence of UV absorption EW on SFR (again derived by inverting the EW vs. metallicity relation) from $z=0-2.2$ galaxies in the same stellar mass range \citep{MANNUCCI10}. The relations are consistent with each other within $1~\sigma$ and indicate a continued relation between SFR and metallicity from low-z to high-z.

Summarizing, we find a significant correlation between UV absorption EW and dust as well as a negative correlation between UV absorption EW and SFR. These relations found here are in good agreement with what is expected from local and low-z galaxies.

\begin{figure}[t!]
\centering
\includegraphics[width=1\columnwidth, angle=0]{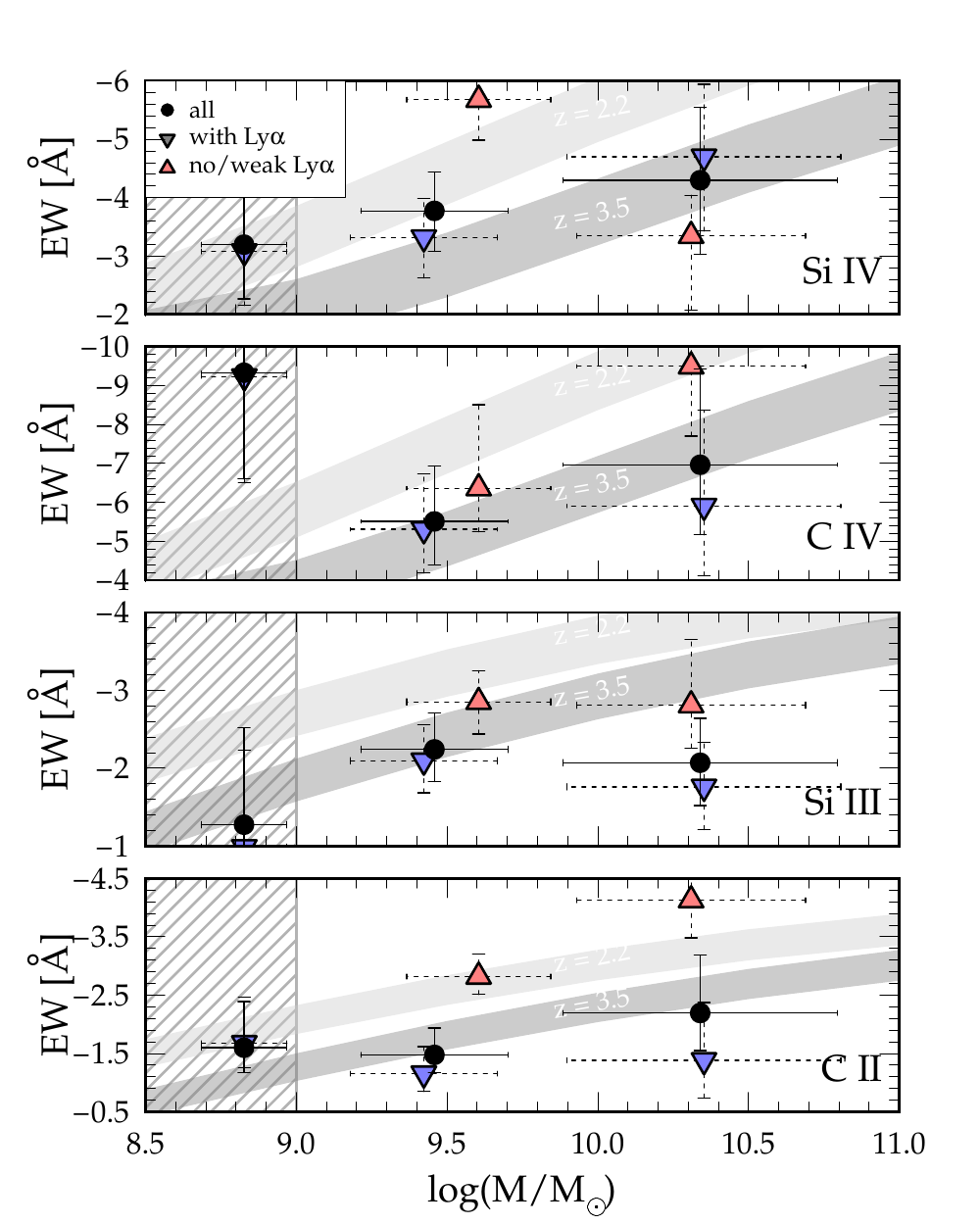}
\caption{Dependence of EW of UV absorption complexes on stellar mass. The points show our data at $z\sim5$ (black dots). We split galaxies with \lya~emission (blue down-triangles) and weak/no \lya~emission (red up-triangles, only show for $\logm>9$ because of their small number). The gray bands show the expected correlation from the mass vs. metallicity relations at $z=2.2$ and $z=3.5$ from the literature. We have converted the metallicity to UV EW by using our relations presented in \autoref{tab:fit}. The data in the hatched region ($\logm<9$) depend significantly on age, especially in the case of \civ~and \siiii~(see text), and are therefore not considered here. At a fixed stellar mass, $z\sim5$ with \lya~emission galaxies show systematically $>1\sigma$ lower UV EWs than $z=2.2$ galaxies, however, similar values than $z=3.5$ galaxies. Galaxies with weak/no \lya~emission show enhanced EW similar to $z\sim2.2$ galaxies.
\label{fig:EWmass}}
\end{figure}

\begin{figure*}[t!]
\centering
\includegraphics[width=1.8\columnwidth, angle=0]{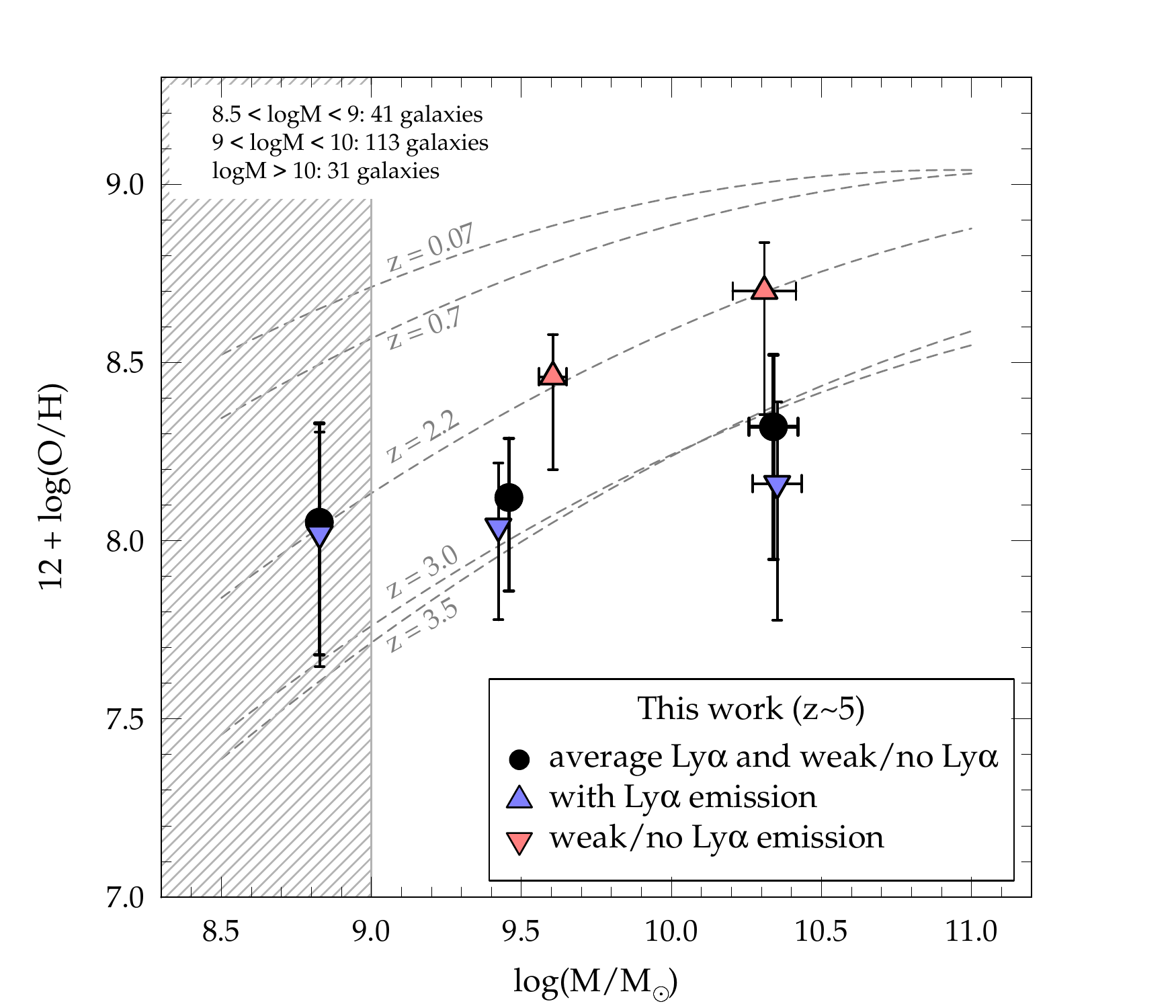}
\caption{
Gas phase metallicity as a function of stellar mass for galaxies at different redshifts. The large symbols show our results at $z\sim5$ split in galaxies with \lya~emission (blue), weak/no \lya~emission (red), and the average of both (black). The metallicities are obtained by a simultaneous fit to all the absorption complexes (see text). The dashed lines show different measurements from the literature at $z=0.07$ \citep{KEWLEY08}, $z=0.7, 2.2, 3.5$ \citep{MAIOLINO08}, and $z=3.0$ \citep{MANNUCCI09}.
\lya~emitting galaxies at $z\sim5$ have a metal content similar to $z\sim3-3.5$ galaxies and only a weak dependence on stellar mass. Galaxies devoid of \lya~emission show a higher metallicity at fixed mass comparable to $z\sim2$ galaxies together with a stronger dependence on stellar mass. This is indicative of them being more evolved systems and highlights the diversity amongst $z\sim5$ galaxies.
\label{fig:mz1}}
\end{figure*}

\subsection{Metallicity vs. stellar mass at $z\sim5$}

We now investigate the dependence of UV absorption EW on stellar mass. For this, we split the sample of our 224 galaxies into three bins of stellar mass: $8.5 < \logm < 9.0$, $9.0 < \logm < 10.0$, and $\logm > 10.0$. We designed these bins to maximize the S/N as well as the baseline in stellar mass, but a different binning does not change our results.

In \autoref{fig:EWmass} (see also \autoref{tab:EWz5}), we show the UV absorption EWs as a function of stellar mass for each of the for absorption complexes. We split galaxies with \lya~emission and weak/no \lya~emission (only show for $\logm>9$ because of their small number). While for $\logm < 9$ galaxies the \siiv~and \cii~is in good agreement with the expectation of an increasing EW with stellar mass, \civ~shows an excess and \siiii~a deficit in EW. This discrepancy can be explained by two reasons, namely \textit{(i)} the lower S/N of the low mass galaxies and \textit{(ii)} their young ages predominantly $< 100~{\rm Myr}$ (also verified by the stellar population ages estimated from SED fitting). As described in \myrefsec{sec:biases}, we correct for the former and therefore this effect cannot explain the discrepancies. The latter is expected to have a dominant impact on the UV EWs. As shown by the BPASS models (see \autoref{fig:EW}), galaxies younger than $\sim100~{\rm Myrs}$ show an excess in \civ~and a deficit in \siiii~compared to older stellar populations at all metallicities. Unfortunately, this trend cannot be verified by the samples used to calibrate the EW vs. metallicity relation since these galaxies are not representative of these young stellar populations.

Focusing on $\logm>9$, we find a systematically higher UV EWs for more massive galaxies in all of the absorption complexes except \siiii. Note that we do not expect all absorption complexes to show the same behavior. This because these elements are produced in different regions of the galaxy. For example, \siiii~may be produced at lower temperature and deeper in the ISM compared to \civ~or \siiv. The expected EW vs. stellar mass relations for $z=2.2$ and $z=3.5$ galaxies derived by inverting our EW vs. metallicity relation are shown as light and dark gray bands. For \lya~emitting galaxies, the EWs are comparable to the ones expected for $z=3-3.5$ galaxies, however, the EW vs. stellar mass relations are shallower for our $z=5$ galaxies by at least $1\sigma$. For galaxies with weak/no \lya~emission, we find higher EWs comparable to galaxies at $z\sim2.2$.

\capstartfalse
\begin{deluxetable*}{l  c cccc | cccc | c}
\tabletypesize{\scriptsize}
\tablecaption{Bias corrected rest-frame EW measurements and estimated metallicities for $z\sim5$ galaxies split in different stellar mass bins.\label{tab:EWz5}}
\tablewidth{0pt}
\tablehead{
\colhead{ } & \colhead{ } & \multicolumn{4}{c}{----------------------- EW [\AA] -----------------------} & \multicolumn{5}{c}{-------------------------- $\oabund$ --------------------------}\\
\colhead{$\logm$} & \colhead{\# galaxies$^{a}$} & \colhead{\siiv} & \colhead{\civ} & \colhead{\siiii} & \colhead{\cii} & \colhead{\siiv} & \colhead{\civ} & \colhead{\siiii} & \colhead{\cii} & \colhead{all$^{b}$} 
}
\startdata
$>8.0$ & 224 (42) & -4.09$^{+0.35}_{-0.45}$  & -6.37$^{+1.26}_{-1.35}$  & -2.20$^{+0.33}_{-0.42}$ & -1.63$^{+0.23}_{-0.28}$  & 8.45$^{+0.13}_{-0.16}$  & 8.31$^{+0.28}_{-0.47}$  & 8.06$^{+0.35}_{-0.32}$  & 8.01$^{+0.21}_{-0.26}$ & 8.30$^{+0.26}_{-0.32}$  \\[0.2cm] 
$>9.0$ & 178 (40) & -3.71$^{+0.55}_{-0.60}$  & -5.54$^{+1.09}_{-1.46}$  & -2.42$^{+0.31}_{-0.36}$ & -1.82$^{+0.36}_{-0.37}$  & 8.35$^{+0.18}_{-0.28}$  & 8.10$^{+0.39}_{-0.86}$  & 8.17$^{+0.35}_{-0.29}$  & 8.07$^{+0.25}_{-0.37}$ & 8.26$^{+0.31}_{-0.51}$  \\[0.2cm] 
$8.5-9.0$ & 41 (2) & -3.19$^{+0.93}_{-0.99}$  & -9.33$^{+2.73}_{-3.05}$  & -1.27$^{+0.67}_{-1.24}$ & -1.60$^{+0.42}_{-0.79}$  & 8.09$^{+0.45}_{-0.73}$  & 8.43$^{+0.25}_{-1.08}$  & 7.54$^{+0.85}_{-0.45}$  & 7.99$^{+0.48}_{-0.53}$ & 8.05$^{+0.56}_{-0.74}$  \\[0.2cm] 
$9.0-10.0$ & 135 (27) & -3.77$^{+0.69}_{-0.67}$  & -5.51$^{+1.12}_{-1.42}$  & -2.24$^{+0.41}_{-0.47}$ & -1.48$^{+0.31}_{-0.46}$  & 8.31$^{+0.24}_{-0.52}$  & 8.14$^{+0.37}_{-0.71}$  & 8.09$^{+0.39}_{-0.42}$  & 7.91$^{+0.29}_{-0.33}$ & 8.12$^{+0.33}_{-0.52}$  \\[0.2cm] 
$>10.0$ & 43 (13) & -4.30$^{+1.27}_{-1.24}$  & -6.97$^{+1.79}_{-2.46}$  & -2.07$^{+0.55}_{-0.57}$ & -2.19$^{+0.64}_{-0.99}$  & 8.37$^{+0.27}_{-0.70}$  & 8.24$^{+0.41}_{-1.02}$  & 7.96$^{+0.42}_{-0.52}$  & 8.13$^{+0.46}_{-0.60}$ & 8.32$^{+0.40}_{-0.74}$ 
\enddata
\tablenotetext{a}{Total numbers of galaxies. The numbers in brackets count galaxies with weak or no \lya emission.}
\tablenotetext{b}{Simultaneous fit to all absorption complexes.}
\end{deluxetable*}
\capstarttrue

Next, we convert these UV EWs into metallicities. For a given EW, we compute its corresponding metallicity by a $\chi^2$ minimization to the 2$^{{\rm nd}}$-order polynomial fit derived in \myrefsec{s:calibration}. The errors are estimated by a Monte-Carlo sampling that accounts for the uncertainty of the EW measurements as well as the input EW vs. metallicity relation. This is done for each single absorption complex. We then compute a combined metallicity for all the absorption complexes by the weighted mean of the normalized probability distribution function derived from the single absorption complexes. The weights are determined from the covariance matrices of the corresponding EW vs. metallicity relations. The results are listed in \autoref{tab:EWz5}.
The symbols in \autoref{fig:mz1} show our results (for galaxies with and without \lya emission) for the combined absorption complexes along with measurements from the literature at $z=0.07$ \citep{KEWLEY08}, $z=0.7, 2.2, 3.5$ \citep{MAIOLINO08}, and $z=3.0$ \citep{MANNUCCI09}. The errors on the points include systematic uncertainties (from the measurement of the EW and the EW vs. metallicity relation) as well as the scatter in the metallicity measurements from single absorption complexes. The data reflects our previous results that \textit{(i)} the metal content of $z=5$ \lya emitting galaxies is more than $2\sigma$ lower than at $z\sim2$ and \textit{(ii)} galaxies with weak/no \lya emission show enhanced metallicities comparable to $z\sim2$ galaxies. The MZ relation of the average population of  $z\sim5$ galaxies is very similar to $z\sim3.5$. However, there are indications that it is slightly shallower.

Summarizing, we find a more than a factor of two lower metal content for the average population of $z=5$ \lya emitting galaxies compared to $z=2$. On the other hand, they show similar metallicities than $z=3-3.5$ galaxies and we find an indication of a shallower MZ relation, however, only with $1\sigma$ significance.

\subsection{\lya~vs. weak/no \lya~emission}

We have designed our sample to be as complete as possible, so it includes galaxies with strong \lya emission as well as galaxies with weak or no emission. In the following, we investigate the properties of these two populations in more detail.
	\autoref{fig:mz1} shows the MZ relation split in galaxies with \lya~emission (blue) and with weak or no emission (red). The latter is only shown at $\logm>9$ due to the lack of a sufficient number of galaxies in the lowest mass bin.
	At a fixed stellar mass, galaxies with weak/no \lya emission show a more than a factor of two higher metal content than galaxies with strong \lya emission at $z\sim 5$. This is also reflected in their larger EWs (see \autoref{fig:EWmass}). In particular, their metal content is similar to that found in $z=2$ galaxies in terms of its amount and also in terms of its dependence on stellar mass, where as strong \lya emitters show no significant MZ relation. \lya is resonantly scattered on neutral gas in the galaxy, which also traces dust. A weak \lya emission therefore occurs in a dusty system, in agreement with having a higher average metal content.
	Summarizing, our results indicate that weak \lya~emitting galaxies are more evolved with higher dust and metal content. This result is strengthened by observation of a less prominent P-Cygni profile in these galaxies (see \autoref{fig:stackall}). Strong \lya~emitting galaxies on the other hand show lower metallicities and a less significant MZ relation, indicative of them being young.


\section{What would we expect the MZ relation to be at high-z?}\label{sec:discussion}

Up to redshifts of $z\sim3$, a strong dependence of metallicity on stellar mass is measured. This indicates the gradual build-up of stellar mass and metallicity over time including feedback, gas inflows and outflows. Furthermore, metallicity is directly linked to the gas and dust content of galaxies \citep[e.g.,][]{ROSEBOOM12,FELDMANN15}, inducing a relation between stellar mass and dust attenuation \citep[e.g.,][]{GARN10,FINKELSTEIN12,SOBRAL12,SANTINI14,HEINIS14,OTEO14}. Also, it is expected that there exists an negative correlation between SFR and metallicity, which might be due to feedback or star-formation fueled by the accretion of pristine gas onto the galaxy \citep[e.g.,][]{MANNUCCI10}. Galaxies at low-z have formed over several billion years and thus had time to reach an equilibrium state between the above quantities and to set up the observed relations. Several studies have shown that these relations hold up to redshifts as high as $z=3-3.5$. A correlation between stellar mass and dust is even expected up to $z=7$ \citep[e.g.,][]{FINKELSTEIN12}, although the stellar mass estimates in these studies are very uncertain due to the lack of deep observed mid-IR data.

\begin{figure}[t!]
\centering
\includegraphics[width=1.0\columnwidth, angle=0]{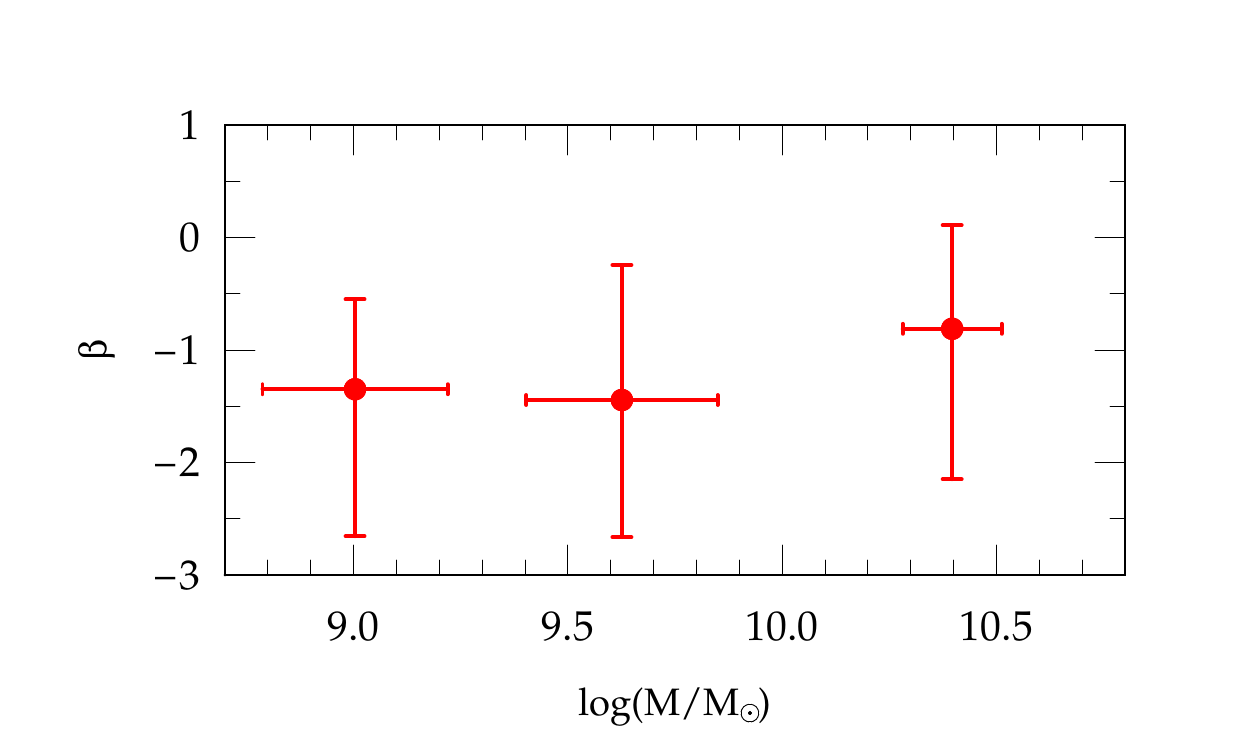}
\caption{Dependence of UV continuum (dust attenuation) on stellar mass for our $z\sim5$ galaxies. We do not find a significant correlation between dust and stellar mass in contradiction to studies at lower redshifts.
\label{fig:betamass}}
\end{figure}

Taking these local relations bluntly at face value and applying them to higher redshifts, we would thus expect positive correlations between metallicity (i.e., UV absorption line EW), stellar mass, and dust extinction as well as a negative correlation between metallicity and SFR. 
Indeed, our data show a correlation between dust attenuation and UV absorption EW as expected from local galaxies, indicating that the dust vs. metallicity relation is intact at $z=5$ (\autoref{fig:EWbeta}). Furthermore, we see a negative correlation between SFR and UV absorption strength consistent with low-z data, indicative of strong accretion of pristine gas in high-z galaxies as fuel for their star-formation (\autoref{fig:EWsfr}). However, we do not see a significant correlation between stellar mass and dust attenuation (\autoref{fig:betamass}) for the average galaxy population and so it is not unexpected that we do not see a significant correlation between stellar mass and metallicity, either (\autoref{fig:mz1}). This indicates that something might be changing at $z\sim5$ and local relations start to weaken.

\textit{What could be the cause for a weak relation between stellar mass and metallicity at $z=5$?}
It might be that our assumption and observational techniques fail at these redshifts.
First, the EW vs. metallicity relation may not be applicable at $z>3$, however, it is unclear why this should happen over this time period. This cannot be tested until JWST. Second, galaxies at high-z are dominated by emission lines that can substantially boost their stellar masses \citep[e.g.,][]{SCHAERER09,ATEK11,STARK13,FAISST15}. Although we include emission lines in our SED templates, their contribution and strength are not known reliably and the obtained stellar masses can still be biased \citep[see e.g.,][]{HSU14}. This effect would in particular boost the masses of low mass galaxies with significant \oiii~emission \citep[][]{ATEK11}. Third, there are indications that the strong line methods (used to calibrate our UV absorption line vs. metallicity relation up to $z=3$) are less reliable at high-z due to the changing internal properties of the galaxies \citep[][]{STEIDEL14,MASTERS14,SHAPLEY15,SANDERS15b}.

Keeping these caveats in mind, there might also be interesting physical reasons responsible for the weakening of the MZ relation.
Several studies indicate that high-z galaxies behave differently and live in different environments than the average galaxy at low-z. They are characterized by a clumpy mode of star-formation \citep[][]{FORSTERSCHREIBER11}, caused by strong gas inflows in the dense and gas-rich high-z universe. They show shorter gas depletion times \citep[e.g.,][]{SCOVILLE15b,SILVERMAN15} and a mass build-up on shorter timescales indicated by their high specific SFRs \citep[][]{STARK13,SPEAGLE14,FAISST15}. All these different properties of high-z galaxies and their surrounding have an impact on the observed MZ relation. In the following, we use a ``bathtub model'' approach \citep[e.g.,][]{LILLY13} to investigate this in more detail.

\begin{figure}[t!]
\centering
\includegraphics[width=1.1\columnwidth, angle=0]{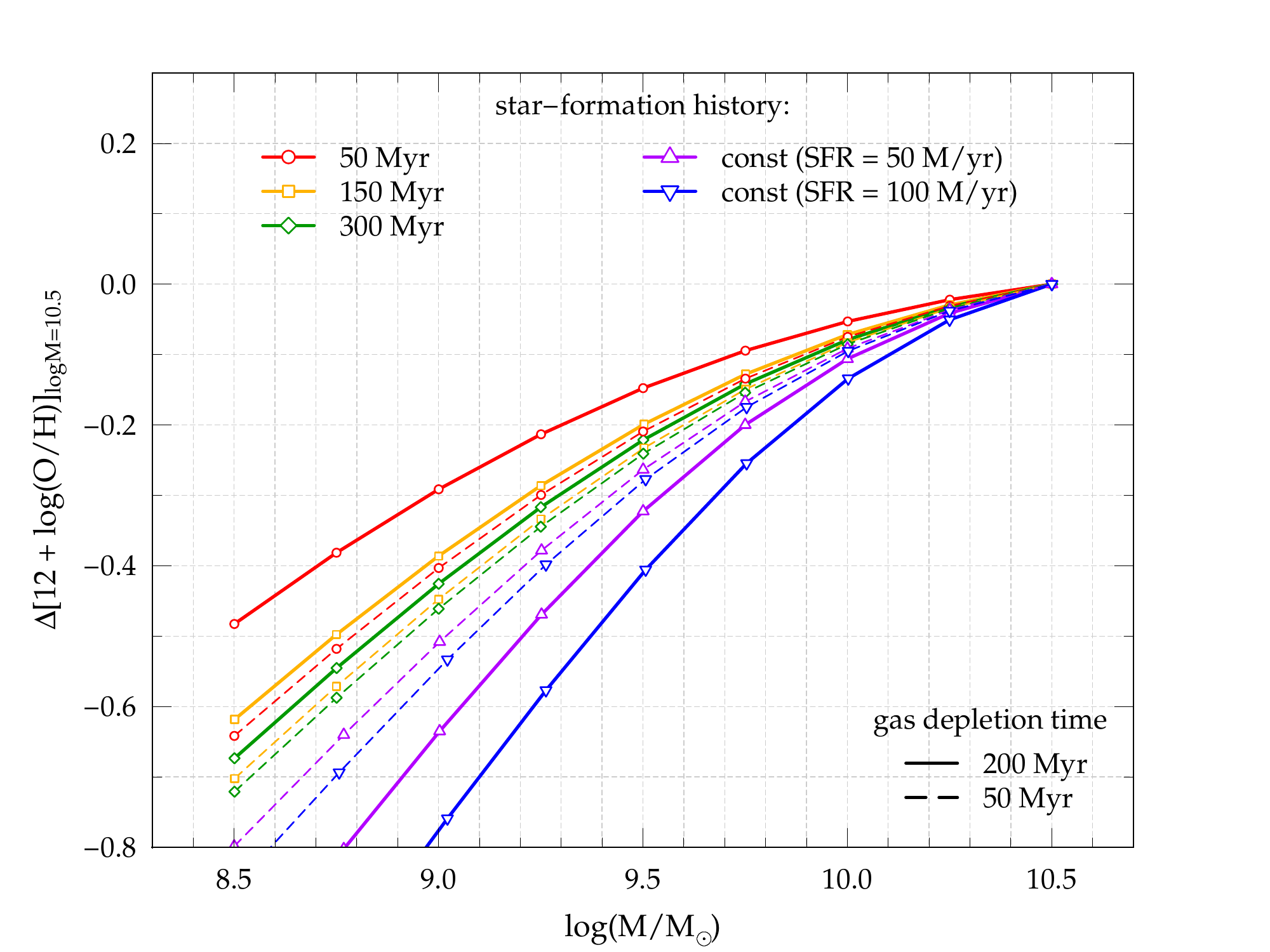}
\caption{Theoretical prediction of the slope of the MZ relation as a function of different SFHs in the ``bathtub'' formalism. The SFH are parametrized by ${\rm SFR}(t) \propto e^{t/\tau}$ with different $\tau$ as well as ${\rm SFR}(t) =~{\rm const}$. We also assume two different gas depletion times (i.e., inverse of star-formation efficiency) of 50 and 200 Myrs \citep[e.g.,][]{SCOVILLE15b}. The MZ relations are normalized to their value at $\logm=10.5$ in order to remove dependences from other constants (e.g., yield).
The slope of the MZ relations significantly change with the assumed SFH. Flatter relations can be achieved by a star-burst like SFH with short $\tau$.
\label{fig:MZslope}}
\end{figure}

In brief, we assume a similar set of differential equation as set up in \citet[][]{FELDMANN15}. We assume a mass dependent mass loading factor as described in their equation~24, which is motivated by hydro-dynamical simulations of galactic winds \citep{HOPKINS12}. For the gas depletion time $t_{{\rm depl}}$ we assume $200~{\rm Myr}$ and $50~{\rm Myr}$ independent of redshift \citep[e.g.,][]{SCOVILLE15b}. We let the galaxies evolve according to two different sets of SFHs in the $\sim1$~billion years to redshift $z=5$. We assume three exponential increasing SFH with $\tau=50$, 150, and 300 Myr as well as a constant SFH with ${\rm SFR}_0=50~{\rm M_{\odot}/yr}$ and ${\rm SFR}_0=100~{\rm M_{\odot}/yr}$. 
\autoref{fig:MZslope} shows how the slope of the MZ relation derived from our simple bathtub model is affected by the different SFHs and gas depletion times. From our simple model, we expect that the slope becomes shallower for star-burst like SFH (i.e., exponentially increasing with a short $\tau$) as well as for short gas depletion times as long as $\tau > t_{{\rm depl}}$.
	\textit{We note, that the uncertainty in our data does not allow us to constrain the model's parameters. Rather, the model should give an idea what could happen at high-z and what we expect in terms of the evolution of the MZ relation to high redshifts.}

\begin{figure}[t!]
\centering
\includegraphics[width=1.1\columnwidth, angle=0]{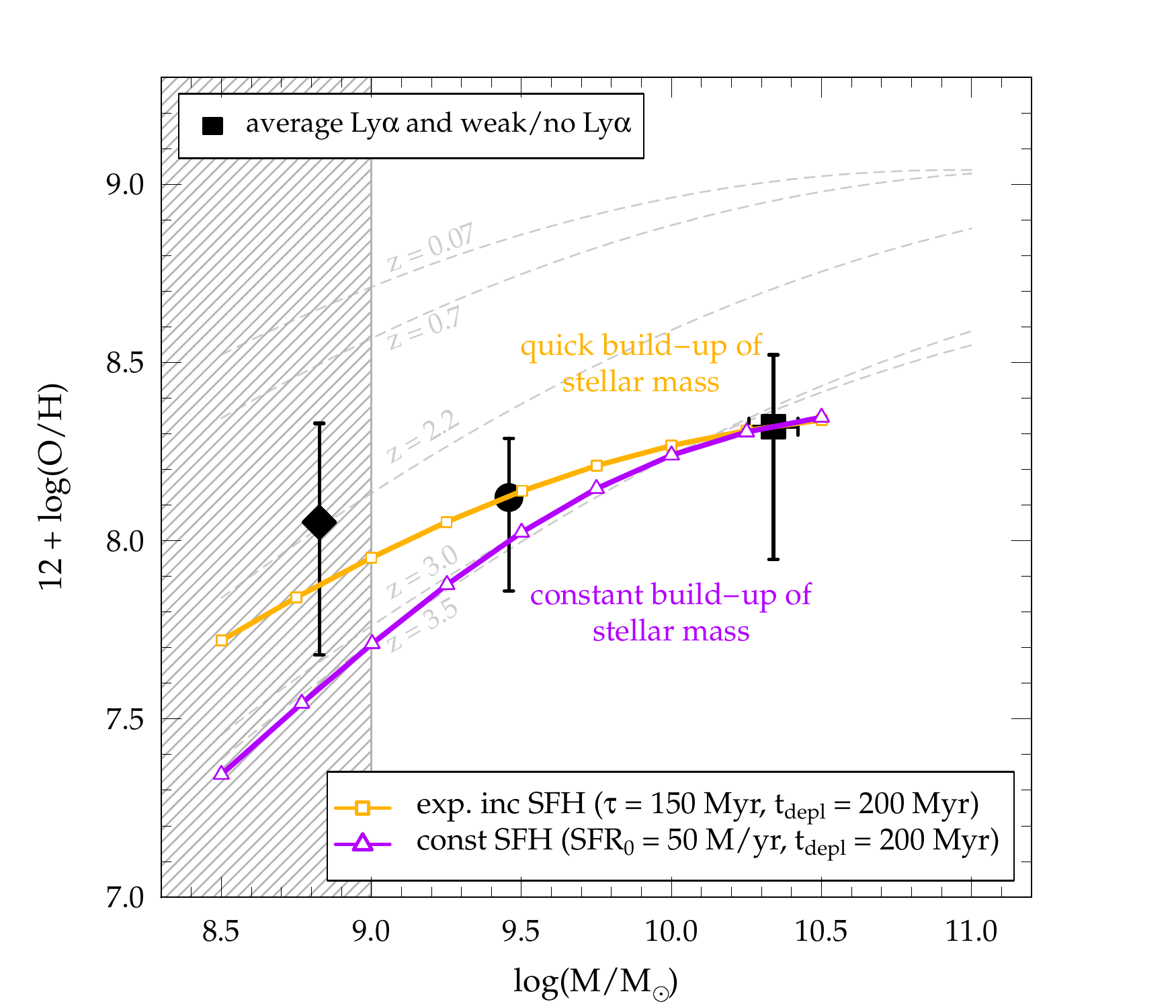}
\caption{
The average MZ relation for our $z\sim5$ galaxies along with two models with exponential increasing ($\tau=150~{\rm Myr}$) and constant (${\rm SFR}=50~{\rm M_{\odot}/yr}$) SFH, normalized to $\logm\sim10.4$ (see \autoref{fig:MZslope}). Our derived MZ relation at $z\sim5$, if real, requires star-formation on short time-scales, while at lower redshifts (e.g., $z\sim3-3.5$) a shallower SFH is favored. This hints towards a rapid growth of massive galaxies in the early epochs of our universe.
\label{fig:mz2}}
\end{figure}

Keeping this in mind, \autoref{fig:mz2} shows a sub-set of models normalized to $\logm=10.5$ along with our estimates at $z=5$ and data at lower redshifts. We pick a constant SFH with ${\rm SFR}_0=50~{\rm Myr}$ and $t_{{\rm depl}}=200~{\rm Myr}$ as well as an exponentially increasing SFH with $\tau=150~{\rm Myr}$ and $t_{{\rm depl}}=200~{\rm Myr}$ matching well the data at $z=5$ and $z=3.5$, respectively. A shallower MZ relation is therefore in agreement with the mass build-up of in high-z galaxies on short timescales as expected from their high specific SFRs. 

A mass build-up on short timescale might also prevent the galaxy population as a whole to have a clear relation between stellar mass and dust obscuration along the line of sight within the galaxy. It indicates that the processes that form the galaxies at high-z are more stochastic than at low-z.
In the same stellar mass range (i.e., without dependence on mass), the dust and metallicity properties of these galaxies vary substantially.
This is probed by our sample consisting of galaxies with strong \lya emission as well as galaxies showing weak/no \lya emission. Since \lya is resonantly scattered off neutral gas (that is correlated with dust), weak \lya emission is indicative of a higher dust column density and therefore a higher metal content of these galaxies. This is in agreement with our findings: weak \lya emitting galaxies show stronger UV absorption (\autoref{fig:stackall}), i.e., a higher metallicity that is similar to $z\sim2$ galaxies (\autoref{fig:mz1}). Furthermore, their less prominent P-Cygni profile is indicative of an older stellar population; hence they are more evolved than strong \lya emitting galaxies. Since they do have the same mass distribution and redshift as galaxies with strong \lya emission, they might have experienced a more gradual formation over a longer time scale, closer to the SFHs of galaxies at $z=2-3$. Interestingly, galaxies with \lya deficit do show a more significant MZ relation (\autoref{fig:mz1}) comparable to $z=2-3$ galaxies. This is in agreement with a SFH with a longer $e-$folding time as shown by our simple bathtub model approach.

Summarizing, we do expect a weakening of the MZ relation at $z=5$ if galaxies are formed on fast, exponential time scales with $e-$folding times of $100-200\,{\rm Myrs}$. 
	The weaker MZ relation indicated by our data is supported by the lack of a significant correlation between stellar mass and dust attenuation. Furthermore, the $z=5$ galaxy population is diverse, including (at the same masses) evolved galaxies with metal contents similar to $z=2$ galaxies. The big diversity might be indicative of the process of galaxy formation at $z=5$ being more stochastic and diverse. This could weaken the clear relations seen at lower redshifts.

\section{Summary \& Conclusions}\label{s:summary}

	Emission features employed to measure metallicity in $z\lesssim3$ galaxies are red shifted out of the wavelength range of current ground-based detectors. It is therefore not possible to probe the metal content of $z>3$ galaxies using these lines until the advent of JWST. Instead, we have to rely on alternative methods. In this paper, we use the correlation between metallicity and the EW of rest-frame UV absorption features to infer the metal properties of a large sample of galaxies at $z\sim5$.

	We compile a sample of $\sim50$ local galaxies and calibrate the relation between metallicity and the most prominent absorption features seen in high-z galaxies (\siiv, \civ, \siiii, \cii). For the first time, we verify this relation to hold up to $z\sim3$ by using a sample of $\sim30$ galaxies at $z=2-3$ with metallicity measurements from strong optical emission lines.
	We then apply (correcting for various biases) this method to a spectroscopic sample of $224$ galaxies at $z\sim5$, which is constructed to be as diverse as possible in terms of physical properties of the galaxies. Most importantly, this sample allows us to investigate directly the properties of galaxies with strong \lya emission as well as no or weak \lya emission.
	
	The findings of this paper are the following.
	
\begin{itemize}

\item The average population of $z\sim5$ galaxies shows very similar gas-phase metallicities as $z\sim3-3.5$ galaxies at a fixed stellar mass but a factor of $\sim2$ lower metal content compared to $z\sim2$.

\item The positive correlation between metallicity and UV continuum slope and the negative correlation between metallicity and SFR agree very well with what is expected from low redshifts. This indicates the dependence of dust and metallicity to hold up to $z\sim5$ as well as a scenario where star-formation is held up by strong inflow of pristine (metal-poor) gas.

\item We do not see a significant correlation between dust attenuation and stellar mass (as seen at lower redshifts). Also, we do not find a significant MZ relation at $z\sim5$ within the uncertainties of our measurements for \lya emitting galaxies.

\item We find a large diversity in our galaxy sample at $z\sim5$. Galaxies with weak/no \lya emission ($\sim 25\%$ at $\logm>10.0$) show a clear MZ relation with metallicities comparable to $z\sim2$. This is indicative of them being more evolved than systems with strong \lya~emission, which is supported by the less prominent P-Cygni profiles in their stacked spectra.

\end{itemize}
	
	Taking our results at face value, there are multiple reasons for the lack of an apparent MZ relation at $z\sim5$. Most of these can be reduced to uncertainties in the measurements of the various parameters (in particular metallicity and stellar mass).
	However, if real, it allows us to constrain the build up of stellar mass in galaxies at these early epochs. Using a ''bathtub-model`` approach, we find that a shallow MZ relation can be caused by a fast build-up of stellar mass on the order of only a couple $100~{\rm Myrs}$.
	The fast formation of these galaxies might cause a more stochastic distribution of the properties and therefore weaken out possible relations that are seen at lower redshifts. This is supported by the observation of our set of galaxies devoid of \lya emission that show a factor of two increased metal content at the same stellar masses than the rest of our sample at $z\sim5$.
		
	For firm conclusions, the reliability of these measurements has to be improved and systematic effects have to be minimized. This can be done by
	\textit{(i)} using stellar population models with a better treatment of the ISM effects to investigate in more depth the age and metallicity, and, 
	\textit{(ii)} targeting specifically low metallicity galaxies at lower redshifts to decrease the systematics of the EW vs. metallicity relation.
	The galaxy sample presented in this work spans a wide range in physical parameters and stems from a spectroscopic survey that is as complete as possible. This sample is therefore ideal to follow-up by JWST, which will be able to measure metallicities for these galaxies and verify our results.

\acknowledgments

	We would like to acknowledge the support of the Keck Observatory staff who made these observations possible as well as Micaela Bagley, Janice Lee, and David Sobral for valuable discussions. AF acknowledges support from the Swiss National Science Foundation.
The authors wish to recognize and acknowledge the very significant cultural role and reverence that the summit of Mauna Kea has always had within the indigenous Hawaiian community. We are most fortunate to have the opportunity to conduct observations from this mountain.


\bibliographystyle{apj}
\bibliography{bibli.bib}

 \appendix
 
\begin{figure}[t!]
\centering
\includegraphics[width=0.8\columnwidth, angle=0]{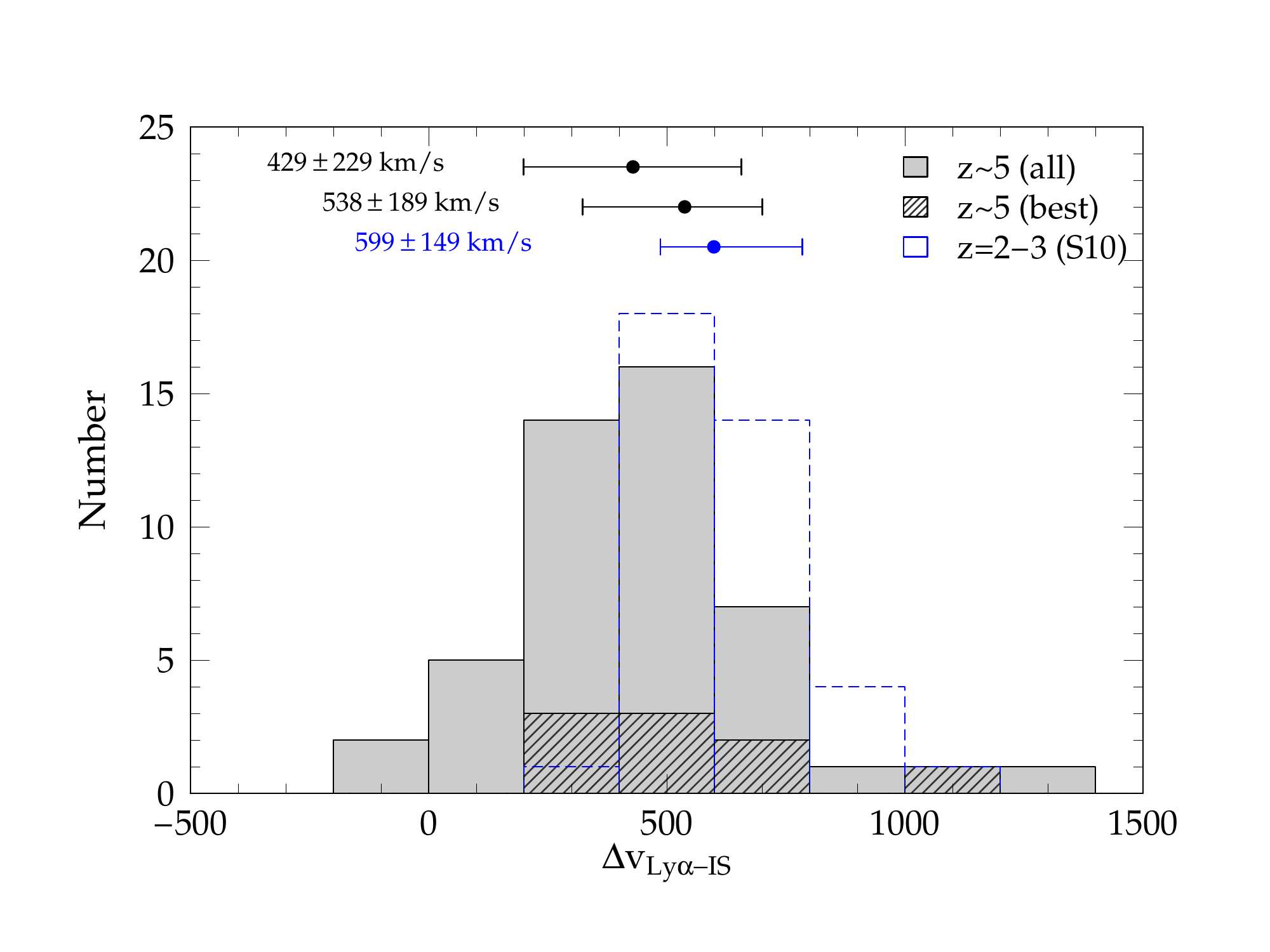}
\caption{Distribution of velocity offset between \lya and IS absorption features for galaxies for which both can be measured (all galaxies: light gray, most secure: dark gray). The data from \citet{STEIDEL10} at $z=2-3$ is shown in dashed blue.
\label{fig:velocities}}
\end{figure}

\section{Velocity offsets due to outflows and winds in $z\sim5$ galaxies}\label{app:velocity}

Winds and outflows are common in local star-burst and star-forming galaxies at high redshifts and related to their metal content. A direct evidence for strong outflows and winds in our sample at $z\sim5$ are the strong P-Cygni profiles seen in the \civ~ and \siiv~absorption lines (see \autoref{fig:stackall}).
	Different velocity offsets of the galaxies can cause biases and uncertainties in the measurements of the spectral properties on stacked spectra. For example, this can lead to a broadening of absorption features or their complete removal in the stacked spectra.
	In this section, we provide estimates of velocity offsets due to outflows and winds in our $z\sim5$ galaxies that we use to quantify the uncertainties in the EW measurements. Moreover, these measurements are useful for further studies of this galaxy sample in the future.

\subsection{Velocity offsets between IS lines and \lya}

	\autoref{fig:velocities} shows the velocity offset of the IS absorption lines with respect to \lya ($\dvlyais$) for individual galaxies for which both \lya and IS absorption is observed. We find a mean velocity offset of $\dvlyais~\sim~538~\pm~189~\kms$ for the most secure IS absorption redshifts and $\dvlyais~\sim~429~\pm 229~\kms$ if including all galaxies with measured IS absorption. This is broadly consistent with measurements of $\sim50$ galaxies at $z\sim2-3$ \citep[][]{STEIDEL10} shown by the the blue dashed histogram.
	The $\dvlyais$ measurements on the composite spectrum for single absorption features are given in \autoref{tab:absfeat} and range between $-1000~\kms$ and $-380~\kms$. The uncertainties of these measurements are conservatively estimated as $\pm100~\kms$, which is twice a resolution element ($100~\kms$ or $0.5~{\rm \AA}$). The median over all absorption lines ($\sim570~\kms$) is consistent with the values for individual galaxies shown in \autoref{fig:velocities}.

\subsection{Systemic velocity offsets of \lya and IS lines}
	
	Finding the absolute velocity components of \lya and the absorption lines ($\Delta v_{\rm sys}$) with respect to systemic is more difficult as we need to measure the systemic redshift of the galaxies from photospheric absorption or emission lines, which is not possible for most of the individual galaxies because of their low S/N.
	However, the complex of photospheric lines at $1300~{\rm \AA}$ consisting of \siiii, \ciii, and \oi~is clearly detected in the composite spectrum. Unfortunately, these lines are blended and therefore an exact systemic redshift cannot be assigned. In addition, because of the different velocity offsets of the individual galaxies, the stacking in the \lya rest-frame changes the measured systemic redshift from the true.
	
	Since our galaxies on average show similar $\dvlyais$ values as the \citet{STEIDEL10} sample at $z=2-3$ (\autoref{fig:velocities}), we use their galaxies (which have systemic redshift from \halpha~measurements) to investigate the systematic biases when measuring $\Delta v_{\rm sys}$ from the composite spectrum. These simulations are detailed in the next section and we summarize the main points in the following.
	First, from \citet{STEIDEL10} we know that the point of largest absorption of the \siiii/\ciii/\oi-complex is a good tracer of the systemic redshift. We then use the systemic redshifts obtained for 8 of our galaxies from the measurement of the \cii~fine-structure line at $157.7 {\rm \mu m}$ using the \textit{Atacama Large Millimeter/Sub-millimeter Array} (ALMA) to verify our approach \citep{CAPAK15}.
	An extensive Monte-Carlo simulation using the \citet{STEIDEL10} sample as input finally shows that there is a bias towards bluer velocity offsets for \lya and IS absorption of $40~\kms$ and $90~\kms$, respectively. This is mainly due to the stacking of galaxies with different $\dvlyais$ in the \lya rest-frame as mentioned above.

	The uncorrected and corrected (in brackets) velocity offsets as well as uncertainties from the Monte-Carlo run are listed in \autoref{tab:absfeat} for the different spectral lines. Note that the velocity offset of the \heii~emission is consistent with the systemic velocity measured by the blended \oi~line; this is an independent verification of the calibration of the systemic rest-frame.
	We find bias corrected velocities for IS absorption features of $-200\pm90~\kms$ and for \lya $340\pm170~\kms$ measured on the composite spectrum.
	Both the \lya and IS absorption velocity offsets are in agreement with studies at $z=2-3$ \citep[$\sim -160~\kms$ and $\sim 450~\kms$, respectively;][]{STEIDEL10} suggesting similar wind properties of these galaxies.

 \subsection{Biases in the measurement of velocity offsets on composite spectrum}

The systemic redshift of individual spectra can not be measured because of their low S/N, on the other hand, several photospheric lines are detected in the composite spectrum (the \siiii/\ciii/\oi~line complex). These are however blended and might be offset from the true due to the stacking of galaxies with different velocity properties.

	We investigate these effects by using a sample of $38$ $z=2-3$ galaxies observed by \citet{STEIDEL10}. These galaxies have UV spectra similar to ours (in particular similar $\dvlyais$) and in addition are observed in \halpha~from which systemic redshift can be measured.	
	First, the analysis by Steidel et al. shows that the \oi~absorption line is a very good tracer of the systemic redshift measured by \halpha~if the deepest absorption of the \siiii/\ciii/\oi~line complex is assigned to \oi. This observation can in addition be tested by a set of 8 galaxies at $z\sim5$ with \lya emission, which are observed with ALMA and have derived systemic redshift from the [\cii] fine structure line at $157.7~{\rm \mu m}$ \citep{CAPAK15}. We stack these 8 galaxies in the \lya rest-frame and measure a velocity offset between \lya and \oi~of $160\pm120~\kms$. This is in excellent agreement with the true \lya velocity offset to systemic (measured by the [\cii] at $157.7~{\rm \mu m}$) which we measure to be $-170~\pm~70~\kms$. The velocity offset between \lya and IS lines is $490 \pm 200~\kms$, consistent with the median of the distribution shown in \autoref{fig:velocities}.
	
	The biases and uncertainties of velocity measurements using the blended \oi~line on the composite spectrum are investigated by a Monte-Carlo simulation.
	We create 38 model galaxies with two different fiducial absorption lines (\oi~for photospheric/systemic line and the redder line of the \siiv~doublet as IS absorption line) as well as \lya based on the velocity offsets shown in figure 2 of \citet{STEIDEL10}. The lines are parametrized by a gaussian with different full-width-at-half-maxima (FWHM) and rest-frame equivalent-width (EW) values. For \lya we choose basis values ${\rm EW}=10~{\rm \AA}$ and ${\rm FWHM}=2~{\rm \AA}$ and for the IS and photospheric lines we choose basis values ${\rm EW}=\left(4,2\right)~{\rm \AA}$ and ${\rm FWHM}=\left(6,8\right)~{\rm \AA}$, respectively. Furthermore, we add random noise to the each of the model spectra such that the S/N is between $2$ and $10$ as for our real galaxies.
	
	We then create 200 composite spectra from the 38 model galaxies, each time changing their EW and FWHM by $20\%$ as well as their noise level (between S/N of 2 and 10).
	On these we measure the velocity offsets of our fiducial \lya and IS absorption line assuming the fiducial \oi~line tracing the systemic redshift.	
	This simulation shows that we are able to recover the input velocity offsets reasonably well within $100~\kms$ from the true. However, we find a systematic bias for both $\dvlya$ and $\dvis$ towards the blue by $40~\kms$ and $90~\kms$, respectively. Also, the uncertainty on the measurements from the Monte-Carlo sampling are on the order of $\pm 180~\kms$ and $\pm 230~\kms$, respectively.
	The corrected velocity values are given in \autoref{tab:absfeat} and are used to estimate the effect of velocity offsets on the measurement of EWs.

\end{document}